\documentclass[bibyear]{aa}
\usepackage{graphicx}
\usepackage{txfonts}
\usepackage{footmisc}

\newcommand{\skc}{{\tt skycorr}}
\newcommand{\Skc}{{\tt Skycorr}}
\newcommand{\xshoot}{\mbox{X-Shooter}}
\begin{document}

   \title{\Skc{}: A general tool for spectroscopic sky subtraction}

   \author{S. Noll\inst{1}
           \and
           W. Kausch\inst{1,2}
           \and
           S. Kimeswenger\inst{3,1}
           \and
           M. Barden\inst{4}
           \and
           A. M. Jones\inst{1}
           \and
           A. Modigliani\inst{5}
           \and
           C. Szyszka\inst{1}
           \and
           J. Taylor\inst{5}
          }

   \institute{Institute for Astro- and Particle Physics, University of
              Innsbruck, Technikerstr. 25/8, 6020 Innsbruck, Austria\\
              \email{stefan.noll@uibk.ac.at}
              \and
              Department of Astrophysics, University of Vienna,
              T\"urkenschanzstrasse 17, 1180 Vienna, Austria
              \and
              Instituto de Astronom\'ia, Universidad Cat\'olica del Norte,
              Avenida Angamos 0610, Antofagasta, Chile
              \and
              International Graduate School of Science and Engineering,
              Technische Universit\"at M\"unchen, Boltzmannstr. 17,
              85748 Garching bei M\"unchen, Germany
              \and
              European Southern Observatory,
              Karl-Schwarzschild-Str. 2,
              85748 Garching bei M\"unchen, Germany
             }

   \date{Received ; accepted }


  \abstract
   {Airglow emission lines, which dominate the optical-to-near-infrared sky
    radiation, show strong, line-dependent variability on time scales from
    minutes to decades. Therefore, the subtraction of the sky background in
    the affected wavelength regime becomes a problem if plain sky spectra have
    to be taken at a different time as the astronomical data.}
   {A solution of this issue is the physically motivated scaling of the airglow
    lines in the plain sky data to fit the sky lines in the object spectrum. We
    have developed a corresponding instrument-independent approach based on
    one-dimensional spectra.}
   {Our code \skc{} separates sky lines and sky/object continuum by an
    iterative approach involving a line finder and airglow line data. The sky
    lines, which mainly belong to OH and O$_2$ bands, are grouped according to
    their expected variability. The line groups in the sky data are then scaled
    to fit the sky in the science data. Required pixel-specific weights for
    overlapping groups are taken from a comprehensive airglow model. Deviations
    in the wavelength calibration are corrected by fitting Chebyshev
    polynomials and rebinning via asymmetric damped sinc kernels. The scaled
    sky lines and the sky continuum are subtracted separately.}
   {ESO-VLT X-Shooter data covering 2.5\,h with a good time resolution were
    selected to illustrate the performance. Data taken six nights and about one
    year before were also used as reference sky data. The variation of the sky
    subtraction quality as a function of time difference between the object and
    sky data depends on changes in the airglow intensity, atmospheric
    transparency, and instrument calibration. Except for short time intervals
    of a few minutes, the sky line residuals were between 2.1 and 5.5 times
    weaker than for sky subtraction without fitting. Further tests show that
    \skc{} performs consistently better than the method of Davies (2007)
    developed for ESO-VLT SINFONI data.}
   {}


   \maketitle


\section{Introduction}\label{sec:intro}

For spectroscopic instrument set-ups and observing programmes that do not
provide plain sky spectra simultaneously with the desired object spectra
(two-dimensional (2D) spectra where the object only covers a small fraction in
spatial direction would be ideal), it is necessary to use sky spectra taken
at a different time as the science spectra for the sky correction. This
especially affects fiber spectrographs, which include instruments using an
integral-field unit (IFU). The strong variability of airglow emission on
various time scales from minutes to decades (see e.g. Khomich et al.
\cite{KHO08}; Patat \cite{PAT08}; Noll et al. \cite{NOL12}) can cause
unavoidable sky correction residua. Significant changes in the airglow emission
can also be expected for angular distances in the order of a few degrees or
even less depending on the characteristics of possible wave patterns, such as
those caused by internal gravity waves (see e.g. Taylor et al. \cite{TAY97}).
Moreover, different telescope positions and ambient conditions may cause
instrument flexures, which result in wavelength shifts of the sky spectrum with
respect to the science spectrum. Hence, the reference sky spectrum has to be
adapted in both, the flux and the wavelength regime, to allow a reasonable sky
background correction.

An illustration of this problem and a solution for the Very Large Telescope
(VLT) near-infrared (near-IR) integral-field spectrograph SINFONI (Eisenhauer
et al. \cite{EIS03}) of the European Southern Observatory (ESO) has been
described by Davies (\cite{DAV07}). His airglow correction method is based on
using arbitrary sky spectra for the sky subtraction in science data by scaling
physically related line groups of the sky spectrum to the corresponding groups
in the science spectrum. Subsequently, the newly created scaled spectrum is
used for the sky correction. Specifically, Davies' method groups emission lines
from vibrational-rotational transitions resulting from non-thermal excitation
processes of the OH molecule (Meinel \cite{MEI50}; Bates \& Nicolet
\cite{BAT50}; Rousselot et al. \cite{ROU00}; Khomich et al. \cite{KHO08}) and
defines wavelength regions where these groups dominate the airglow emission.
These wavelength ranges are scaled in the sky spectrum to match the
corresponding flux in the object spectrum. This approach only works for the
line component of the sky emission. Since object and sky continua cannot
satisfyingly be separated, the sky continuum cannot be adapted and has to be
subtracted before the scaling procedure for the line emission can start.
Davies' code considers a greybody for the thermal continuum, which is fitted to
the data. Other continuum sources that dominate the emission at wavelengths
shorter than the $K$ band (e.g. airglow (pseudo-)continua, scattered moonlight,
zodiacal light, and instrument-related continua; see Noll et al. \cite{NOL12})
are not taken into account. In the $J$ and $H$ bands covered by the SINFONI
data, these contributions are distinctly fainter than the airglow lines. Since
Davies' method is restricted to OH airglow lines beyond 1\,$\mu$m\footnote{The
version 2.0 of Davies' code also allows simple scaling of the strong O$_2$ band
at 1.27\,$\mu$m. This feature is not described in Davies (\cite{DAV07}).}, it
can only be applied to near-IR wavelength regions where other line emission is
negligible. Davies' method has successfully been implemented for the ESO
SINFONI data reduction pipeline (Modigliani et al. \cite{MOD07}).

In this paper, we will present the instrument-independent code
\skc{}\footnote{The code can be downloaded from web pages located at ESO
({\tt http://www.eso.org/pipelines/skytools/}) as well as the University of
Innsbruck ({\tt http://www.uibk.ac.at/eso/software/}).} for the sky correction
based on one-dimensional (1D) spectra. It was inspired by Davies' pioneering
method. In comparison, \skc{} allows a more accurate subtraction of airglow
lines due to its more detailed airglow model (see Noll et al. \cite{NOL12} for
the optical range), the pixel-specific scaling of the lines, and an advanced
approach for separating lines and continuum. Moreover, the code can also be
applied to optical wavelengths. Section~\ref{sec:method} describes the
algorithm implemented in \skc{}. A test data set for illustrating the
performance is presented in Sect.~\ref{sec:data}. The test results are
discussed in Sect.~\ref{sec:evaluation}. This also involves a comparison with
Davies' method. Finally, Sect.~\ref{sec:conclusions} provides a summary and
contains our conclusions.

\section{Method and use}\label{sec:method}

\begin{figure}
\centering
\includegraphics[width=8.8cm,clip=true]{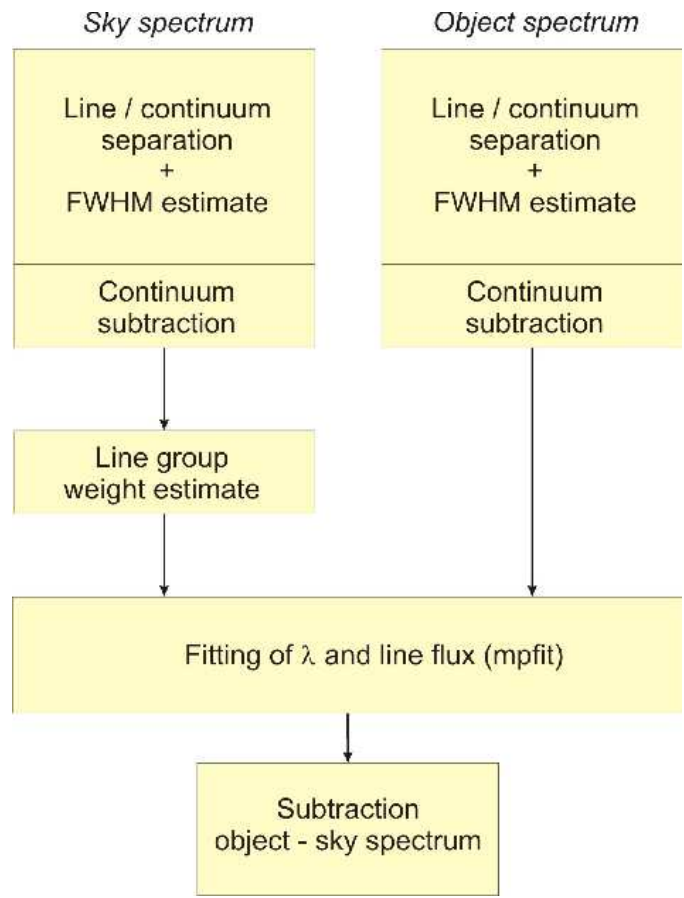}
\caption[]{Overview of the \skc{} workflow}
\label{fig:overview}
\end{figure}

\begin{table}
\caption[]{\Skc{} default parameter set-up}
\label{tab:setup}
\centering
\small
\vspace{5pt}
\begin{tabular}{@{\,}l l l@{\,}}
\hline\hline
\noalign{\smallskip}
Parameter$^\mathrm{a}$ & Value & Short description \\
\noalign{\smallskip}
\hline
\noalign{\smallskip}
{\sc vac\_air} & vac & wavelengths in vacuum or air \\
{\sc fwhm} & 5.0 & initial estimate of line FWHM (in pixels) \\
{\sc varfwhm} & 0 & linear increase of line width with \\
& & wavelength? (1 = yes, 0 = no) \\
{\sc ltol} & 1e-2 & relative FWHM convergence criterion \\
{\sc min\_line\_dist} & 2.5 & minimum distance to neighbouring lines \\
& & for isolated lines (in FWHM) \\
{\sc fluxlim} & -1 & minimum line peak flux (in median flux \\
& & of identified lines) \\
{\sc ftol} & 1e-3 & relative $\chi^2$ convergence criterion \\
{\sc xtol} & 1e-3 & relative parameter convergence criterion \\
{\sc wtol} & 1e-3 & convergence criterion for iterative \\
& & improvement of wavelength grid \\
{\sc cheby\_max} & 7 & maximum degree of polynomial for \\
& & wavelength grid correction \\
{\sc cheby\_min} & 3 & minimum degree of polynomial for \\
& & wavelength grid correction \\
{\sc cheby\_const} & 0. & initial constant term for wavelength grid \\
& & correction \\
{\sc rebintype} & 1 & type of rebinning (0 = simple, \\
& & 1 = asymmetric, damped sinc kernel) \\
{\sc weightlim} & 0.67 & minimum relative weight of the strongest \\
& & line group of a pixel \\
{\sc siglim} & 15. & $\sigma$ limit for excluding outliers \\
{\sc fitlim} & 0. & lower relative uncertainty limit for fitting \\
& & a line group \\
\noalign{\smallskip}
\hline
\end{tabular}
\begin{list}{}{}
\item[$^\mathrm{a}$] Input/output parameters are not listed.
\end{list}
\end{table}

In the following, we describe the \skc{} sky correction procedure in detail.
The algorithm is sketched in Fig.~\ref{fig:overview}. Its main purpose is the
removal of sky emission lines in a science spectrum by means of a scaled
reference sky spectrum. Any continuum component in the input science and
reference sky spectra has to be removed in advance. Hence, as a first step,
pixels belonging to lines and continuum have to be identified, separated, and
masked accordingly. The line identification is discussed in
Sect.~\ref{sec:linesearch}. This procedure also derives the typical line Full
Width Half Maximum (FWHM, Sect.~\ref{sec:FWHM}). The next step is a fit of the
continuum pixels only, which is subtracted from the input spectrum to obtain a
continuum-free spectrum (Sect.~\ref{sec:contsub}). In the third step, a weight
mask for the different airglow line groups in the reference sky spectrum is
calculated incorporating an extended version of the sky radiance model
described in Noll et al. (\cite{NOL12}) (Sect.~\ref{sec:airglow}). It is an
essential input for scaling the reference sky line spectrum to fit the science
line spectrum in the subsequent step (Sect.~\ref{sec:linefit}). The fitting
procedure also allows an optimisation of the wavelength grids
(Sect.~\ref{sec:wavegrid}). The final step of the sky correction procedure is
the sky subtraction itself. This operation is just the subtraction of the
best-fit sky line spectrum and the unscaled sky continuum spectrum from the
input science spectrum.

\Skc{} has been designed to be instrument independent. It only requires a
1D science and reference sky spectrum as input. Both should be taken with the
same instrumental set-up and reduced in the same way. For a good sky
subtraction (especially for the unscaled continuum), both spectra should refer
to a similar integrated sky area. For slit spectrographs, this depends on the
slit width and the extraction in spatial direction. The former is also
important for the spectral resolution. For data without flux calibration, the
science and sky spectra should correspond to the same exposure time. It is
expected that both files have the same format. \Skc{} accepts ASCII tables,
FITS tables, and 1D FITS images. The corrected output files will have the
format of the input files. Several additional files are written to the output
directory, which contain results of the algorithm and can be used to evaluate
the quality of the sky subtraction. \Skc{} has also been designed to minimise
user interaction. Nevertheless, there might be code parameters that could be
modified to improve the sky subtraction results. The relevant parameters of the
\skc{} driver file and their default values are listed in
Table~\ref{tab:setup}. They are discussed in the following subsections. More
details on the file structures and code parameters can be found in the \skc{}
User Manual, which is provided along with the code.

\subsection{Line finder}\label{sec:linesearch}

\begin{figure}
\centering
\includegraphics[width=8.8cm,clip=true]{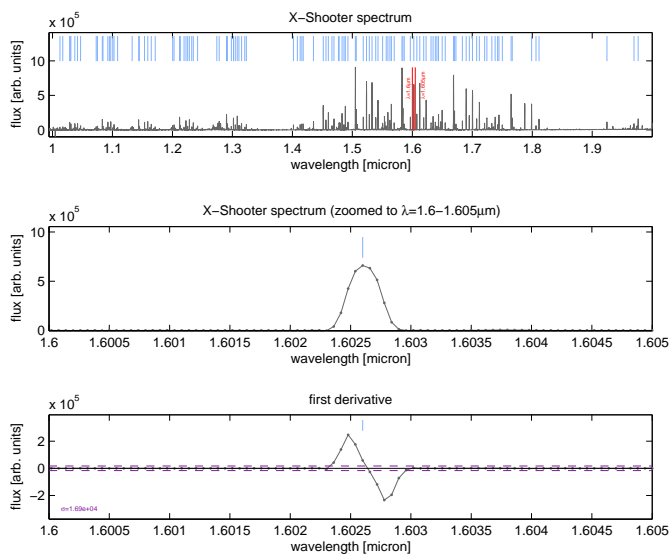}
\caption[]{Line detection in an X-Shooter sky spectrum. While the upper panel
shows a wide wavelength range, the middle panel focuses on a narrow range
containing a prominent single emission line. In the lower panel, the first
derivative of this line is given. Such patterns are used to identify emission
lines. The resulting detections are marked by light blue vertical lines in the
panels.}
\label{fig:linesearch}
\end{figure}

Spectral lines are identified by an approach that uses the first derivative of
the spectrum (see Fig.~\ref{fig:linesearch}). Thus, line pixels can be
recognised by their large flux gradients. Emission line peaks can be identified
by a change from positive to negative values of the first derivative. The
approach is insensitive to slowly varying continua. All detected lines are
assembled in a line list, which is refined by an iterative method depending on
the estimation of the sky line FWHM (see Sect.~\ref{sec:FWHM}). This procedure
requires the identification of strong, isolated lines. Therefore, the line
finder checks the previously identified line peaks, applying criteria that
reject spurious detections and line blends. Lines that are sufficiently
separated from other lines and whose peaks have a symmetric shape are marked as
isolated. The former criterion can be influenced by the user. The parameter
file includes the unitless scaling parameter {\sc min\_line\_dist} (see
Table~\ref{tab:setup}) that is internally multiplied by the line FWHM in
pixels, where a first guess value is also included in the configuration file.

For strong and well separated lines, the described line finding method is very
robust. However, for low resolution spectra with many blended lines, a major
fraction of lines may be missing. Therefore, the line list of the airglow model
is used (see Sect.~\ref{sec:airglow}) to find previously unidentified sky lines
that have fluxes above the median flux of the lines identified by the
derivative approach times the value of the {\sc fluxlim} parameter specified in
the parameter file. By default, {\sc fluxlim} is set to -1. In this case, sky
lines are identified by successive iterations, starting with a threshold level
of 0.005 and doubling this value in a subsequent iteration if the limit does
not result in sufficient continuum pixels for the continuum interpolation (see
Sect.~\ref{sec:contsub}), i.e. at least 20\% of all pixels distributed over
more than 90\% of the wavelength range. The iterative doubling of the threshold
is repeated until the criterion is fulfilled or a maximum value of 0.08 is
reached. The number of pixels characterised as line pixels for each line
included from the line list depends on the derived line FWHM (see
Sect.~\ref{sec:FWHM}). The combination of line pixels identified by both
methods gives a good estimate of the spectral ranges covered by significant
airglow lines.

\subsection{Line FWHM estimator}\label{sec:FWHM}

For computing the airglow model (see Sect.~\ref{sec:airglow}), it is required
to convert total line fluxes as provided by the input line list into fluxes per
wavelength unit. Consequently, it is necessary to know the typical FWHM of the
airglow lines. The line finder described in the previous
Sect.~\ref{sec:linesearch} searches for isolated lines that are suitable for
deriving a line width. After the subtraction of the continuum (see
Sect.~\ref{sec:contsub}), the FWHM estimation can be performed by fitting a
Gaussian to the line pixels belonging to each isolated line. The assumed shape
of the line profile is not critical for the required accuracy. The FWHM
measurements of all isolated lines are averaged to obtain the typical FWHM of
the input spectrum. In order to avoid blended lines contributing to the
resulting mean, a $\sigma$-clipping approach is applied to skip suspiciously
high FWHM. If less than five isolated lines remain after clipping, the median
FWHM is taken.

Since the FWHM estimation and the search for isolated lines (see
Sect.~\ref{sec:linesearch}) affect each other, an iterative approach is
required, where the line finder, continuum subtractor, and FWHM estimator are
called in a loop in order to obtain a stable and trustworthy FWHM. This
iterative procedure is terminated if convergence is reached for the mean FWHM.
The convergence criterion is provided by the parameter {\sc ltol} (see
Table~\ref{tab:setup}).

For spectrographs whose spectra show a roughly linear increase of the FWHM with
wavelength and cover a wide wavelength range (see Sect.~\ref{sec:approach}),
one can set the parameter {\sc varfwhm} to 1. In this case, the FWHM estimates
of the individual lines are converted to correspond to the FWHM which would be
measured at the central wavelength of the full spectrum, assuming a linear
change of the FWHM with wavelength. The converted FWHM are then used to
calculate the mean FWHM as discussed above. A linear change of the FWHM is
also assumed for the separation of lines and continuum (see
Sect.~\ref{sec:linesearch}) and the determination of the pixel contributions of
the different line groups (see Sect.~\ref{sec:linefit}).

\subsection{Continuum subtraction}\label{sec:contsub}

\begin{figure}
\centering
\includegraphics[width=8.8cm,clip=true]{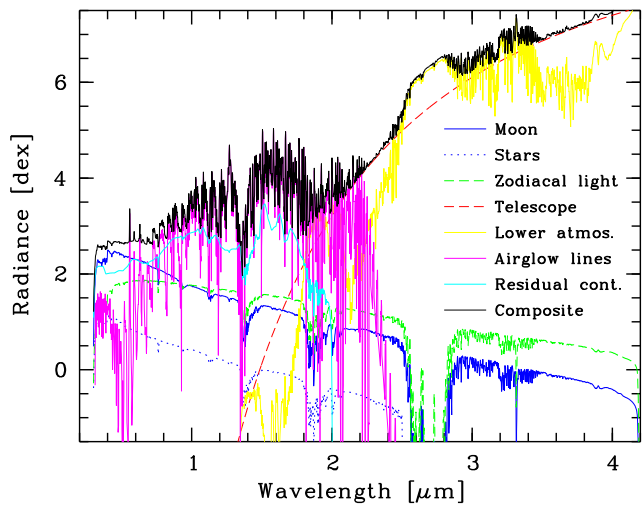}
\caption[]{Components of the Cerro Paranal sky model for wavelengths between
$0.3$ and 6\,$\mu$m in logarithmic radiance units. The example with Moon above
the horizon shows the scattered moonlight, scattered starlight, zodiacal light,
thermal emission by telescope and instrument, molecular emission of the lower
atmosphere, airglow emission lines of the upper atmosphere, and
airglow/residual continuum. The (optical) model components are described in
Noll et al. (\cite{NOL12}) except for the improved scattered moonlight model,
which is discussed in Jones et al. (\cite{JON13}).}
\label{fig:logcomp}
\end{figure}

Apart from spectra of astronomical objects, plain sky spectra also show
continuum emission. The main components are scattered moonlight, scattered
starlight, zodiacal light, thermal emission from the lower atmosphere by
greenhouse gases and the telescope itself, unresolved airglow bands, and
airglow continuum emission, which is related most probably to chemical
reactions in the upper atmosphere involving nitric oxide (Khomich et
al. \cite{KHO08} and references therein; Noll et al. \cite{NOL12}). In
addition, an instrument-related continuum due to internal scattering and
similar effects could significantly contribute (see e.g. Ellis \&
Bland-Hawthorn \cite{ELL08}; Vernet et al. \cite{VER11}). As
Fig.~\ref{fig:logcomp} indicates, the main continuum component is the
airglow/residual continuum\footnote{Note that the airglow continuum is very
difficult to determine. Its measured intensity strongly depends on the accuracy
of the other components, the quality of the flux calibration, and possible
instrumental continua. For this reason, this model component should also be
seen as residual continuum.}, which dominates shortwards of the thermal regime
with the exception of the ultraviolet (UV) and optical if the Moon is up.

The variable airglow continuum (see Noll et al. \cite{NOL12}) and the thermal
continuum, which can significantly change by temperature differences of a few
degrees, cannot be corrected by a fitting procedure like for the airglow lines
(see Sect.~\ref{sec:airglow}), since object and sky continuum cannot be
separated in the science spectrum (cf. Sect.~\ref{sec:intro}). Hence, \skc{}
performs a simple continuum subtraction without scaling. For reliable results,
this approach requires that the object continuum is distinctly brighter than
the variation in the sky continuum (see Sect.~\ref{sec:sampprop} for more
details). Note that the strong thermal methane and water vapour lines of the
lower atmosphere at wavelengths beyond 2.3\,$\mu$m are also handled as
continuum. This can lead to significant sky subtraction residuals for
inaccurate wavelength solutions (see Sect.~\ref{sec:wavegrid}).

\Skc{} obtains the continua in the input science and sky spectra using line
identification flags set in the course of the line search described in
Sect.~\ref{sec:linesearch}. All pixels not flagged as line pixels are connected
by linear interpolation. If the identification of continuum pixels is reliable,
this is the most efficient approach even in the case of line blends covering
wide wavelength ranges.

\subsection{Airglow model}\label{sec:airglow}

\begin{table}
\caption[]{Description of $A$~groups in the input line list}
\label{tab:Agroups}
\centering
\footnotesize
\vspace{5pt}
\begin{tabular}{c c c l}
\hline\hline
\noalign{\smallskip}
ID & $N_\mathrm{lin}$ & $\lambda$ range & Description \\
   &                 & [$\mu$m]        &             \\
\noalign{\smallskip}
\hline
\noalign{\smallskip}
 1 &  61 & 0.314 - 0.872 & green O\,I at 557.7\,nm \\
   &     &               & + unidentified lines \\
 2 &   3 & 0.589 - 0.770 & Na\,I\,D + other lines \\
   &     &               & from alkali metals \\
 3 &  23 & 0.389 - 0.845 & red O\,I at 630.0\,nm \\
   &     &               & + other thermospheric lines \\
 4 &   1 & 0.467 - 0.467 & OH(7-0) \\
 5 &   8 & 0.491 - 0.495 & OH(8-1) \\
 6 &  22 & 0.519 - 0.536 & OH(9-2) \\
 7 &  12 & 0.526 - 0.535 & OH(6-0) \\
 8 &  23 & 0.554 - 0.570 & OH(7-1) \\
 9 &  41 & 0.587 - 0.634 & OH(8-2) \\
10 &  49 & 0.624 - 0.655 & OH(9-3) \\
11 &   2 & 0.672 - 0.674 & OH(10-4) \\
12 &  27 & 0.614 - 0.695 & OH(5-0) \\
13 &  83 & 0.647 - 0.754 & OH(6-1) \\
14 & 113 & 0.681 - 0.782 & OH(7-2) \\
15 & 111 & 0.720 - 0.815 & OH(8-3) \\
16 &  72 & 0.768 - 0.822 & OH(9-4) \\
17 &   7 & 0.827 - 0.839 & OH(10-5) \\
18 &  85 & 0.745 - 0.910 & OH(4-0) \\
19 & 113 & 0.781 - 0.914 & OH(5-1) \\
20 & 111 & 0.826 - 0.916 & OH(6-2) \\
21 & 110 & 0.873 - 0.937 & OH(7-3) \\
22 & 116 & 0.931 - 1.007 & OH(8-4) \\
23 & 120 & 0.994 - 1.081 & OH(9-5) \\
24 & 100 & 0.965 - 1.043 & OH(3-0) \\
25 & 110 & 1.015 - 1.098 & OH(4-1) \\
26 & 112 & 1.069 - 1.168 & OH(5-2) \\
27 & 118 & 1.129 - 1.236 & OH(6-3) \\
28 & 120 & 1.197 - 1.314 & OH(7-4) \\
29 & 124 & 1.275 - 1.420 & OH(8-5) \\
30 & 128 & 1.366 - 1.531 & OH(9-6) \\
31 & 112 & 1.392 - 1.558 & OH(2-0) \\
32 & 118 & 1.461 - 1.654 & OH(3-1) \\
33 & 122 & 1.537 - 1.743 & OH(4-2) \\
34 & 122 & 1.622 - 1.842 & OH(5-3) \\
35 & 124 & 1.717 - 1.978 & OH(6-4) \\
36 & 126 & 1.825 - 2.110 & OH(7-5) \\
37 & 130 & 1.951 - 2.265 & OH(8-6) \\
38 & 130 & 2.101 - 2.454 & OH(9-7) \\
39 & 450 & 0.314 - 0.532 & O$_2$(A-X) (Herzberg I) \\
40 &   5 & 0.324 - 0.410 & O$_2$(c-X) (Herzberg II) \\
41 & 396 & 0.326 - 0.550 & O$_2$(A'-a) (Chamberlain) \\
42 &  65 & 0.382 - 0.509 & O$_2$(c-b) \\
43 & 208 & 0.656 - 0.806 & O$_2$(b-X) (v' > v'') \\
44 & 194 & 0.761 - 0.816 & O$_2$(b-X) (v' = v'') \\
45 & 103 & 0.861 - 0.922 & O$_2$(b-X) (v' < v''; \\
   &     &               & including atm. 0-1 band) \\
46 & 161 & 1.240 - 1.305 & O$_2$(a-X)(0-0) \\
   &     &               & (IR atmospheric system) \\
47 &  73 & 1.555 - 1.598 & O$_2$(a-X)(0-1) \\
   &     &               & (IR atmospheric system) \\
\noalign{\smallskip}
\hline
\end{tabular}
\end{table}

\begin{table}
\caption[]{Description of $B$~groups in the input line list}
\label{tab:Bgroups}
\centering
\footnotesize
\vspace{5pt}
\begin{tabular}{c c l l}
\hline\hline
\noalign{\smallskip}
ID & Molecule & Upper state(s) & Remarks$^\mathrm{a}$ \\
\noalign{\smallskip}
\hline
\noalign{\smallskip}
 1 & OH & $X^2\Pi_{1/2}$, $J' = 1/2$  & Q2(0.5), P2(1.5) \\
 2 & OH & $X^2\Pi_{3/2}$, $J' = 3/2$  & Q1(1.5), P1(2.5) \\
 3 & OH & $X^2\Pi_{1/2}$, $J' = 3/2$  & R2(0.5), Q2(1.5), P2(2.5) \\
 4 & OH & $X^2\Pi_{3/2}$, $J' = 5/2$  & R1(1.5), Q1(2.5), P1(3.5) \\
 5 & OH & $X^2\Pi_{1/2}$, $J' = 5/2$  & R2(1.5), Q2(2.5), P2(3.5) \\
 6 & OH & $X^2\Pi_{3/2}$, $J' = 7/2$  & R1(2.5), Q1(3.5), P1(4.5) \\
 7 & OH & $X^2\Pi_{1/2}$, $J' = 7/2$  & R2(2.5), Q2(3.5), P2(4.5) \\
 8 & OH & $X^2\Pi_{3/2}$, $J' = 9/2$  & R1(3.5), Q1(4.5), P1(5.5) \\
 9 & OH & $X^2\Pi_{1/2}$, $J' = 9/2$  & R2(3.5), Q2(4.5), P2(5.5) \\
10 & OH & $X^2\Pi_{3/2}$, $J' = 11/2$ & R1(4.5), Q1(5.5), P1(6.5) \\
11 & O$_2$ & $b^1\Sigma^+_g$, $J' = 0, 2, 4$ & \\
12 & O$_2$ & $b^1\Sigma^+_g$, $J' = 6, 8$ & \\
13 & O$_2$ & $b^1\Sigma^+_g$, $J' = 10, 12$ & \\
14 & O$_2$ & $b^1\Sigma^+_g$, $J' = 14, 16$ & \\
15 & O$_2$ & $a^1\Delta_g$, $J' = 2, 4$ & $v'' = 0$ \\
16 & O$_2$ & $a^1\Delta_g$, $J' = 6, 8$ & $v'' = 0$ \\
17 & O$_2$ & $a^1\Delta_g$, $J' = 10, 12$ & $v'' = 0$ \\
18 & O$_2$ & $a^1\Delta_g$, $J' = 14, 16$ & $v'' = 0$ \\
19 & O$_2$ & $a^1\Delta_g$, $J' = 18, 20$ & $v'' = 0$ \\
20 & O$_2$ & $a^1\Delta_g$, $J' = 20, 22$ & $v'' = 0$ \\
21 & O$_2$ & $a^1\Delta_g$, $J' = 2, 4$ & $v'' \ne 0$ \\
22 & O$_2$ & $a^1\Delta_g$, $J' = 6, 8$ & $v'' \ne 0$ \\
23 & O$_2$ & $a^1\Delta_g$, $J' = 10, 12$ & $v'' \ne 0$ \\
24 & O$_2$ & $a^1\Delta_g$, $J' = 14, 16$ & $v'' \ne 0$ \\
\noalign{\smallskip}
\hline
\end{tabular}
\footnotesize
\begin{list}{}{}
\item[$^\mathrm{a}$] OH rotational transitions (identified by the branch and
total angular momentum of the lower state $J''$) or lower vibrational level
$v''$ for \mbox{O$_2$(a-X)} transitions
\end{list}
\end{table}

\begin{figure}
\centering
\includegraphics[width=8.8cm,clip=true]{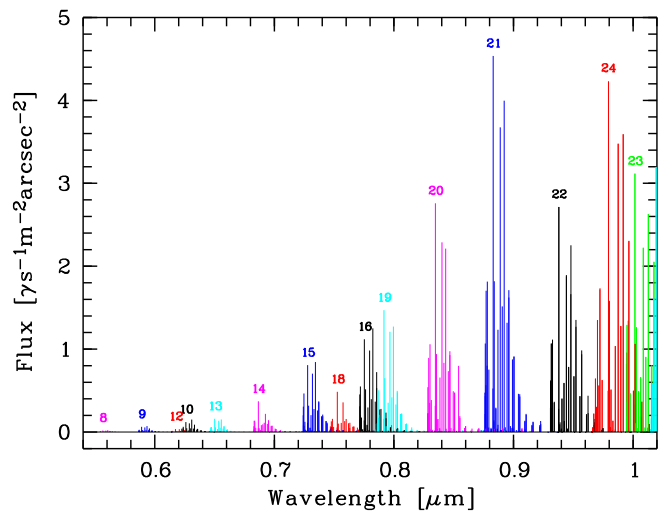}
\caption[]{$A$-group identifications of OH bands (cf. Table~\ref{tab:Agroups})
in the wavelength range between 0.54 and 1.02\,$\mu$m that have Q1(1.5) lines
(see Table~\ref{tab:Bgroups} and Fig.~\ref{fig:ohband}) stronger than
0.01\,$\gamma\,{\rm s}^{-1}\,{\rm m}^{-2}\,{\rm arcsec}^{-2}$. The wavelengths and
zenithal mean fluxes (considering absorption in the lower atmosphere) tabulated
in the input line list are plotted. Note that the bands with numbers up to 21
appear twice as strong as the bands at longer wavelengths, since the
corresponding lines were taken from the Hanuschik (\cite{HAN03}) atlas, where
OH doublets are often unresolved and are listed as one line only.}
\label{fig:ohbands_opt}
\end{figure}

\begin{figure}
\centering
\includegraphics[width=8.8cm,clip=true]{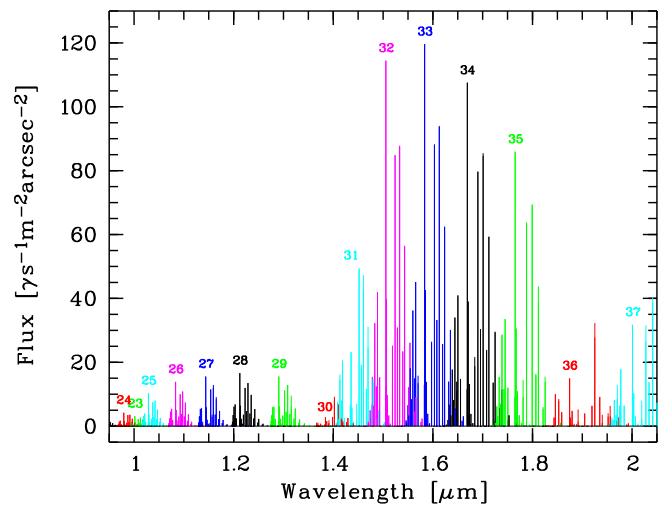}
\caption[]{$A$-group identifications of the OH bands (cf.
Table~\ref{tab:Agroups}) in the wavelength range between 0.95 and 2.05\,$\mu$m.
The wavelengths and zenithal mean fluxes tabulated in the input line list are
plotted.}
\label{fig:ohbands_ir}
\end{figure}

\begin{figure}
\centering
\includegraphics[width=8.8cm,clip=true]{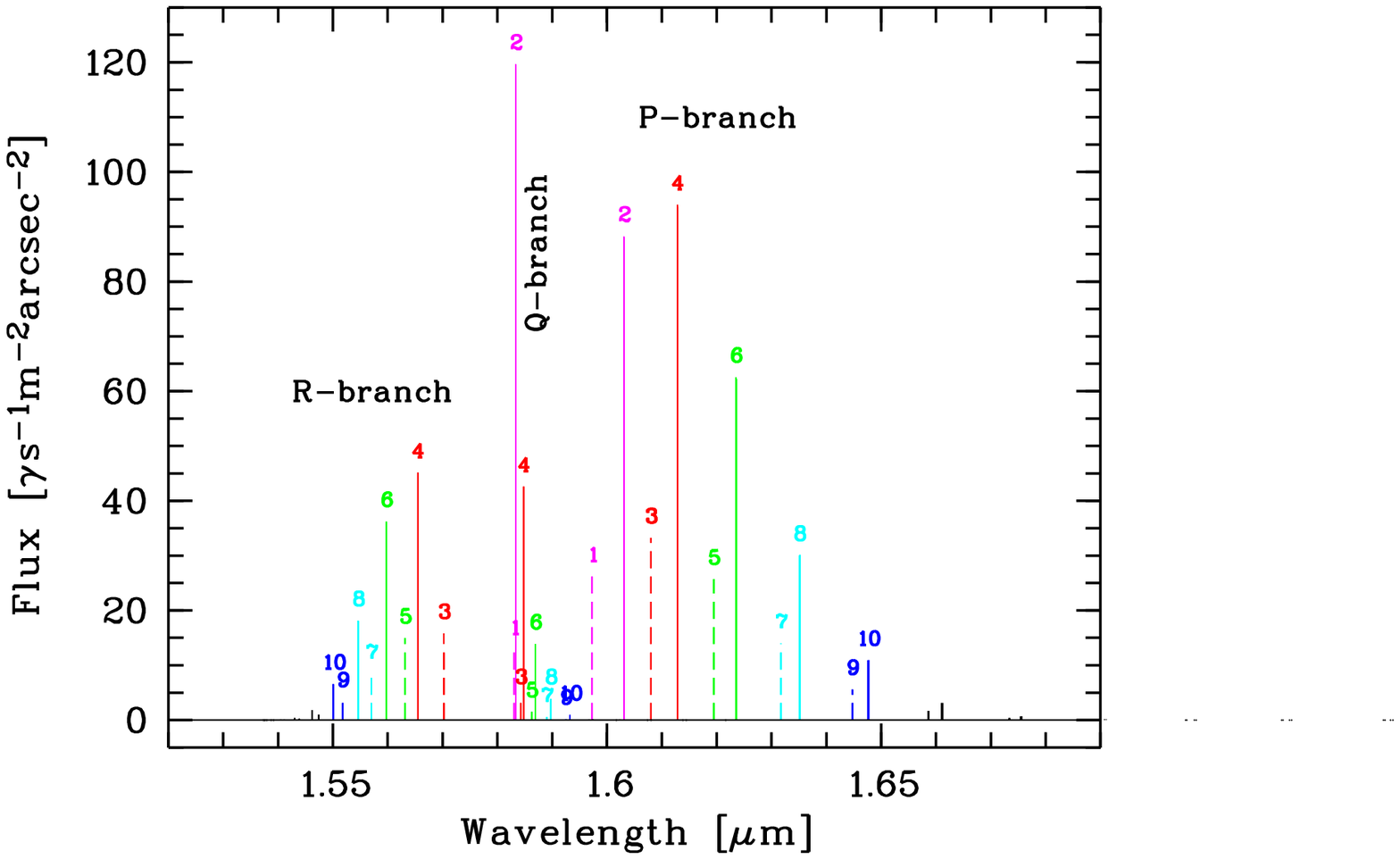}
\caption[]{$B$-group identifications of the transitions of an OH band with the
same rotational upper state (cf. Table~\ref{tab:Bgroups}). The tabulated
wavelengths and zenithal mean fluxes of the lines of the \mbox{OH(4-2)} band
are shown as example. Dashed and solid lines indicate transitions of the
$X^2\Pi_{1/2}$ and $X^2\Pi_{3/2}$ state, respectively. The figure also indicates
the R-, Q-, and P-branches that correspond to transitions with a change of the
total angular momentum by -1, 0, and 1, respectively.}
\label{fig:ohband}
\end{figure}

\begin{figure}
\centering
\includegraphics[width=8.8cm,clip=true]{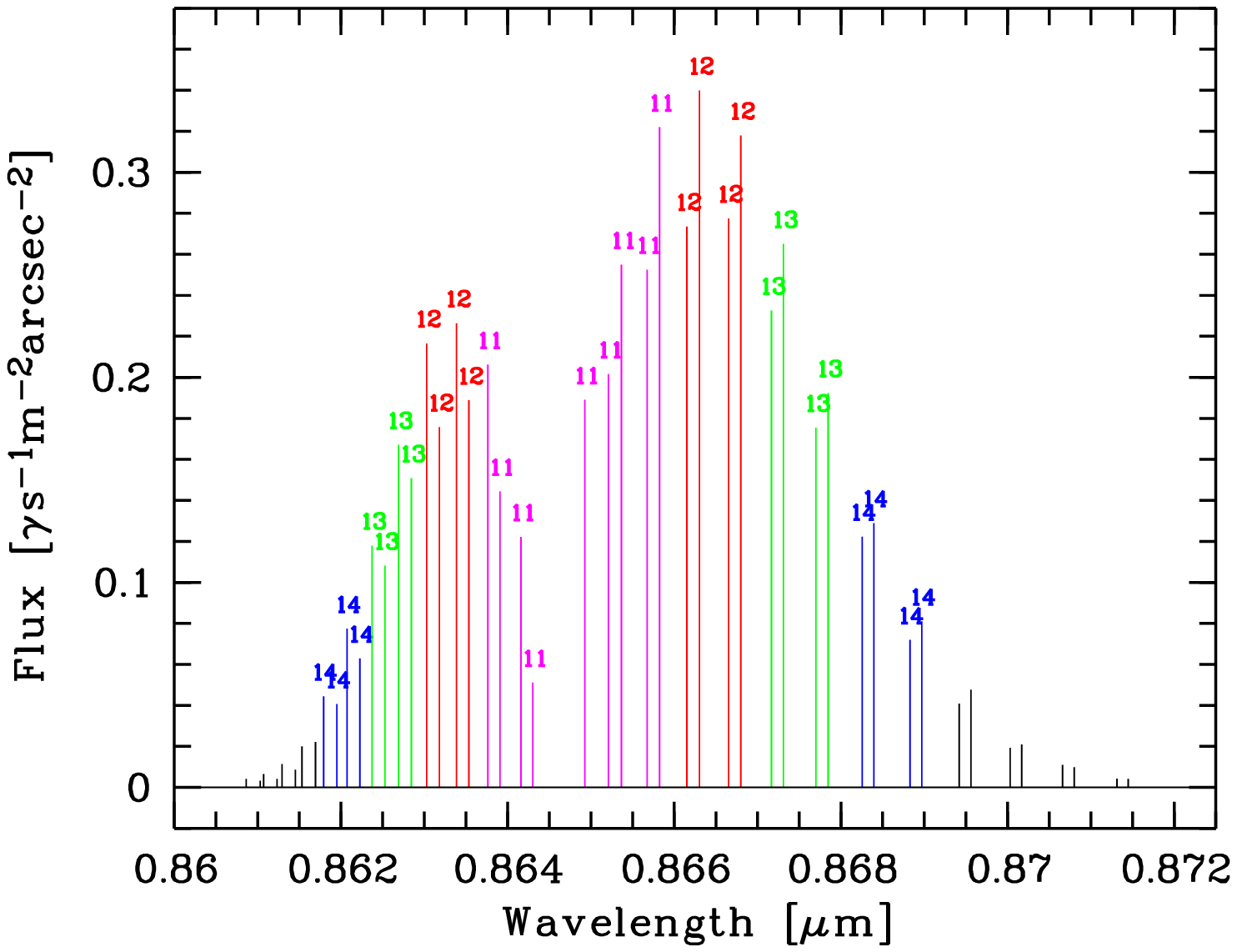}
\caption[]{$B$-group identifications of the transitions of the band
\mbox{O$_2$(b-X)(0-1)} with a similar rotational upper state (cf.
Table~\ref{tab:Bgroups}). The tabulated wavelengths and zenithal mean fluxes of
the lines of the 4 different branches (2 R- and 2 P-branches) are shown (cf.
Fig.~\ref{fig:ohband}).}
\label{fig:o2b01band}
\end{figure}

\begin{figure}
\centering
\includegraphics[width=8.8cm,clip=true]{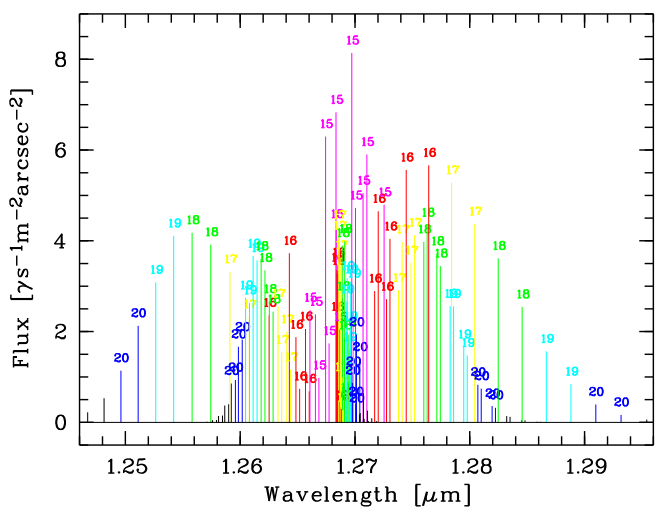}
\caption[]{$B$-group identifications of the transitions of the band
\mbox{O$_2$(a-X)(0-0)} with a similar rotational upper state (cf.
Table~\ref{tab:Bgroups}). The tabulated wavelengths and zenithal mean fluxes of
the lines of the 9 different branches (3 R-, 3 Q-, and 3 P-branches) are shown
(cf. Fig.~\ref{fig:ohband}). The band is strongly affected by self absorption
in the lower atmosphere.}
\label{fig:o2a00band}
\end{figure}

\begin{figure}
\centering
\includegraphics[width=8.8cm,clip=true]{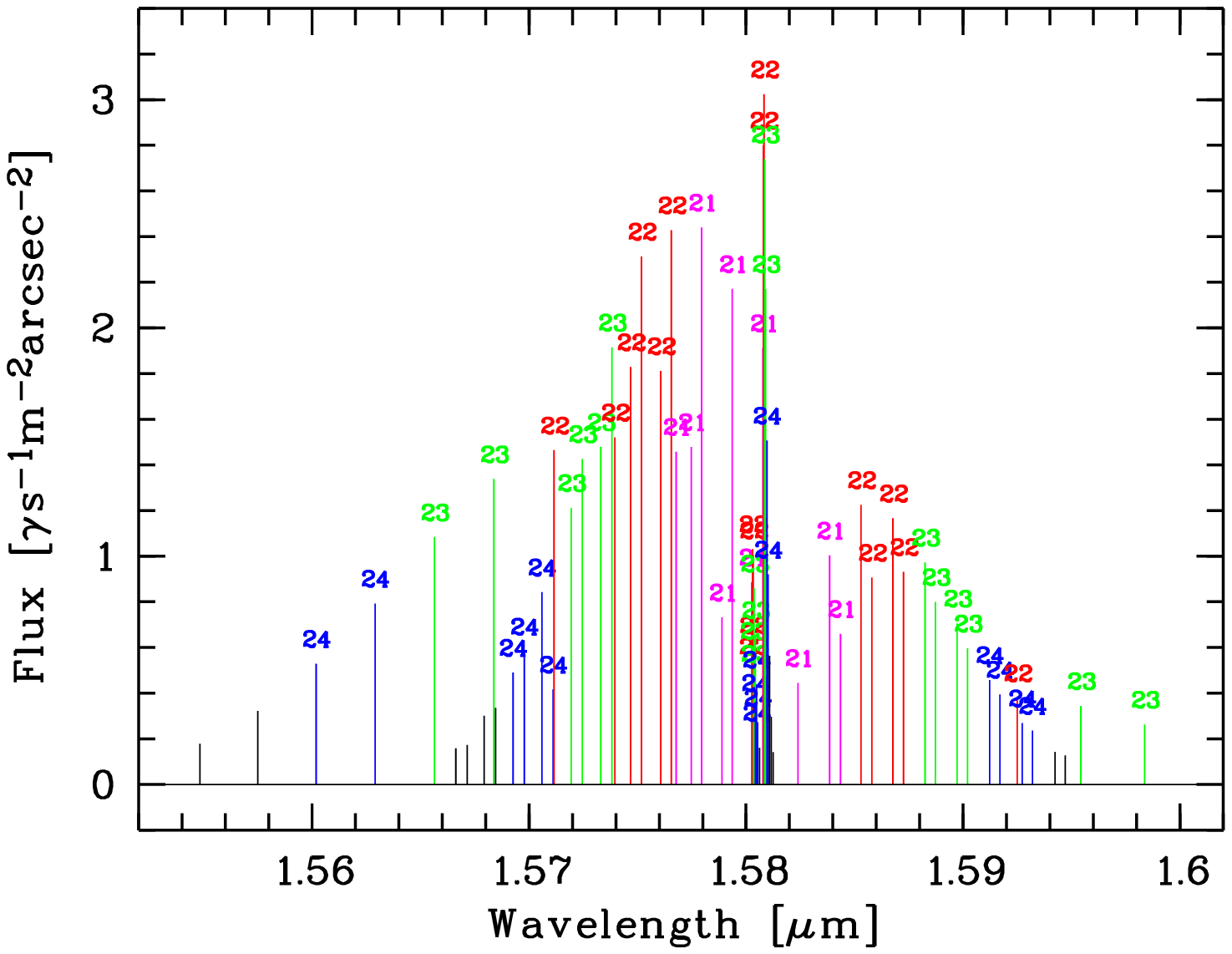}
\caption[]{$B$-group identifications of the transitions of the band
\mbox{O$_2$(a-X)(0-1)} with a similar rotational upper state (cf.
Table~\ref{tab:Bgroups}). The tabulated wavelengths and zenithal mean fluxes of
the lines of the 9 different branches (3 R-, 3 Q-, and 3 P-branches) are shown
(cf. Fig.~\ref{fig:ohband}).}
\label{fig:o2a01band}
\end{figure}

The wavelength range from the near-UV to the near-IR is characterised by strong
emission lines. Most of them constitute band structures. This airglow (see
Khomich et al. (\cite{KHO08}) for a comprehensive discussion) mostly originates
in the mesopause region at about 90\,km. In addition, some lines arise in the
ionospheric F2-layer at about 270\,km. In general, airglow is caused by
chemiluminescence, i.e. chemical reactions that lead to light emission by the
decay of excited states of reaction products. Apart from atomic oxygen and
sodium, the oxygen (O$_2$) and hydroxyl (OH) molecules are the most important
reaction products in this context. In general, airglow lines show strong
variability from time scales in the order of minutes to decades. This behaviour
can be explained by the solar activity cycle, seasonal changes in temperature,
pressure, and chemical composition of the emission layers, the day-night
contrast, dynamical effects such as planetary and gravity waves, or geomagnetic
disturbances. The dynamical effects also cause spatial intensity variations. In
addition, the airglow intensity depends on the projected emission layer width,
which is a function of the zenith distance (van Rhijn \cite{RHI21}; Noll et al.
\cite{NOL12}).

As the airglow is highly variable, the strength of the emission lines in a
reference sky spectrum for sky correction has to be adapted. This is achieved
by a fitting procedure that is discussed in Sect.~\ref{sec:linefit}. Since
object emission lines should not be reproduced by the optimised sky spectrum
and finally removed, it is advisable to scale as many lines as possible by a
single fitting parameter. Every group should contain airglow lines that are not
affected by object lines and that can be used to determine a realistic
correction factor for the reference sky spectrum. The number of lines that can
be combined is limited by the fact that they should show an almost identical
variability behaviour.

As a basis for the definition of suitable line groups, the airglow line model
developed for the Cerro Paranal sky radiance model (Noll et al. \cite{NOL12})
was used. This semi-empirical model consists of a line list with line
intensities for mean observing conditions and prescriptions for the correction
of the line strength depending on molecular species, solar activity, season,
time of night, and zenith distance of the target. The latter three input
parameters can be retrieved from the FITS header of the sky spectrum file. The
solar activity is traced by the solar radio flux at 10.7\,cm, which can be
provided either by the parameter {\sc solflux} in the configuration file or by
the corresponding monthly average (default) in a file offered by
{\tt www.spaceweather.gc.ca}. In the wavelength range from $0.3143$ to
$0.9228$\,$\mu$m, the line list consists of data taken from Cosby et al.
(\cite{COS06}). They incorporated the emission line atlas of Hanuschik
(\cite{HAN03}) based on observations with the VLT high-resolution echelle
spectrograph UVES (Dekker et al. \cite{DEK00}). In order to fill a gap at about
$0.86$\,$\mu$m, this list was supplemented by unpublished UVES data of R.
Hanuschik. At longer wavelengths, the calculated OH lines of Rousselot et al.
(\cite{ROU00}) were included. However, their line strengths were corrected for
the Einstein factors of Goldman et al. (\cite{GOL98}) instead of using the
original ones of Mies (\cite{MIE74}). This resulted in correction factors for
OH band strengths between 0.38 and 2.06. Moreover, the flux decrease of airglow
lines by molecular absorption in the lower atmosphere was corrected. To this
end, Gaussian profiles for each airglow line with Doppler line widths for
typical temperatures of about 200\,K were convolved with the high-resolution
($\lambda / \Delta\lambda \approx 10^6$) Cerro Paranal annual-mean transmission
curve for an airmass of 1.25\footnote{Although the atmospheric transmission
depends on airmass and weather conditions, only a fixed airglow flux correction
was applied in order to avoid time-consuming calculations at very high
resolution and the input of temperature and water vapour profiles. Moreover,
the optical airglow atlas of Hanuschik (\cite{HAN03}) is also characterised by
a fixed transmission correction due to the use of UVES mean spectra. For most
observing conditions, the deviation of the true airglow absorption from the
assumed one is expected to be minor in terms of the results of \skc{}.} (see
Noll et al. \cite{NOL12}). The transmission curve was computed by means of the
radiative transfer code LBLRTM (see Clough et al. \cite{CLO05}). Note that
\skc{} applies the Cerro Paranal mean transmission curve for zenith (corrected
for the target airmass) to the unextincted fluxes in the input line list. For
this purpose, the line list contains separate columns for unextincted line
fluxes and zenithal transmission values. Finally, the Rousselot et al. lines
were scaled to the Cosby et al. lines between $0.642$ and
$0.858$\,$\mu$m\footnote{The calculated OH line spectrum of Rousselot et al.
(\cite{ROU00}) covers a wider wavelength range ($0.614 - 2.624$\,$\mu$m) than
the published atlas related to observations with the VLT spectrograph ISAAC
($0.997 - 2.252$\,$\mu$m).}. The strongest O$_2$ bands in the near-IR at $1.27$
and $1.58$\,$\mu$m were included in the line list by adding data from the
HITRAN molecular line database (see Rothman et al. \cite{ROT09}). The mean band
strength was roughly estimated evaluating the ratio of O$_2$ to OH lines in 26
NIR-arm \xshoot{} spectra (see Vernet et al. \cite{VER11} for details on the
instrument; see also Sect.~\ref{sec:data}). For this purpose, the O$_2$ lines
had to be extincted depending on the airmass values of the \xshoot{} spectra.
In the case of the $1.27$\,$\mu$m band, this caused significant changes in the
line fluxes due to the strong resonant absorption of airglow photons by
tropospheric/stratospheric O$_2$ molecules in the ground state.

The Cerro Paranal sky model assigns the listed lines to the five different
variability classes (1) green O\,I, (2) Na\,I\,D, (3) red O\,I, (4) OH, and (5)
O$_2$ (see Noll et al. \cite{NOL12}). These variability classes result
from analysing a sample of 1189 optical FORS1 spectra (Patat \cite{PAT08}).
From this sample, the lines' dependence on solar radio flux and time of
observation was derived. The latter was quantified using a grid of six double
month periods starting with Dec/Jan and three night time bins of equal length.
The reference line strengths in the line list represent the mean of the five
solar activity cycles 19 to 23, i.e. the years 1954 to 2007. The ratios of line
strengths of different variability classes can easily vary by a factor of two
and more.

Assigning airglow lines to the five classes using the rough predictions from
the sky model is not sufficient for achieving a line intensity accuracy on the
per cent level, which is required for a sky subtraction procedure like \skc{}.
Typically, intensity variations of lines within a variability class are
stronger than can be tolerated.

In principle, an identical variability behaviour can be expected for
transitions with the same upper state. In this case, the ratios of line
intensities should be fixed and only determined by quantities such as Einstein
coefficients and statistical weights. On the other hand, the excitation and
population of different energy levels depends on variable quantities such as
temperature and chemical abundances. Therefore, it is promising to define line
groups depending on the upper energy level. However, taking all relevant states
of the molecules OH and O$_2$ into account would result in a very large number
of line groups. Moreover, each group would consist of only a few significant
lines. This would result in statistical fluctuations, which could make the line
intensity correction uncertain if crucial lines of a group were affected by
e.g. detector defects or object emission lines (see Sect.~\ref{sec:linefit}).
Fortunately, as their energies are rather different, it is possible to separate
electronic, vibrational, and rotational transitions of molecules. The
electronic/vibrational transition determines the band and the rotational
transition identifies a single line or doublet (as in case of OH) within a
band. Since the distribution of energy levels is very similar for all bands of
an electronic transition, each line can be assigned to two different classes
that are defined by the upper vibrational and rotational state. This approach
reduces the number of required line groups significantly. Moreover, for OH only
the electronic ground state is relevant, which splits up into the sub-levels
$X^2\Pi_{1/2}$ and $X^2\Pi_{3/2}$ due to the coupling of spin and orbital angular
momentum (see Rousselot et al. \cite{ROU00}). For O$_2$, the electronic
transitions are more important than the vibrational ones, since the intensity
differences of the bands of an electronic transition are very large.
Consequently, there are only three O$_2$ bands that significantly contribute to
the airglow, namely \mbox{O$_2$(b-X)(0-1)}\footnote{The notation used is as
follows: molecule (upper $-$ lower electronic state) (upper $-$ lower
vibrational state). The letters `a', `b', and `X' are shortcuts for the states
$a^1\Delta_g$, $b^1\Sigma^+_g$, and $X^3\Sigma^-_g$. The vibrational states are
numbered depending on the energy and starting from 0 for the lowest level.}
(the very strong \mbox{(0-0)} band is almost completely absorbed in the lower
atmosphere), \mbox{O$_2$(a-X)(0-0)}, and \mbox{O$_2$(a-X)(0-1)} (see Khomich et
al. \cite{KHO08}). The O$_2$ bands at near-UV and blue wavelengths are weak and
can only be resolved at very high resolution. For most instruments, these bands
therefore appear as a pseudo continuum.

Tables~\ref{tab:Agroups} and \ref{tab:Bgroups} list the final grouping of
airglow lines. Line groups with the same upper electronic/vibrational level are
called ``$A$~groups'' and those with the same (OH) or a similar (O$_2$) upper
rotational level are labelled as ``$B$~groups''. Most OH bands (apart from a
few very weak ones) are identified in Figs.~\ref{fig:ohbands_opt} and
\ref{fig:ohbands_ir}. Although bands such as \mbox{OH(4-1)} and \mbox{OH(4-2)}
have the same upper vibrational level, they represent independent variability
groups. Since real data suffer from calibration uncertainties, this procedure
is necessary, even though the number of $A$~groups increases significantly. As
OH bands with the same upper vibrational level are widely separated, it is
safer to vary such bands independently. Figures~\ref{fig:ohband},
\ref{fig:o2b01band}, \ref{fig:o2a00band}, and \ref{fig:o2a01band} show
identifications of the rotational $B$~groups for OH(4-2) as an example of an
OH band, \mbox{O$_2$(b-X)(0-1)}, \mbox{O$_2$(a-X)(0-0)}, and
\mbox{O$_2$(a-X)(0-1)}, respectively. Although the two \mbox{O$_2$(a-X)} bands
belong to the same roto-vibrational system, their $B$~groups were defined
separately due to the completely different line flux distribution. This is
caused by self absorption in the \mbox{(0-0)} band. $B$~groups of O$_2$ bands
consist of lines from two rotational upper levels in order to make sure that
enough lines can be identified for the group scaling (see
Sect.~\ref{sec:linefit}). The weak lines of each band are not included in a
$B$~group, since they are difficult to fit. Furthermore, this measure avoids a
degeneration of fit parameters.

The described grouping is more complex than the one of Davies (\cite{DAV07}),
since Davies only incorporates near-IR OH bands, \mbox{O$_2$(a-X)(0-0)}, and
two rotational groups resembling our $B\,2$ and $B\,4$ classes (see
Table~\ref{tab:Bgroups}).

\subsection{Airglow line fitter}\label{sec:linefit}

To prepare a reference sky line spectrum for the sky subtraction in a science
spectrum, the line fluxes have to be adapted. To this end, Davies
(\cite{DAV07}) divides the wavelength range into sections depending on the OH
band structure and subsequently scales these segments independently according
to the sections' flux ratio of the science and sky spectra. This can be
problematic where different line groups have significant overlap. While OH
bands in $H$-band spectra do not significantly overlap and are not blended with
strong bands of other molecules, at lower wavelengths the situation is less
favourable (see Sect.~\ref{sec:airglow}). Even so, it is difficult to measure
the scaling factors for groups with the same upper rotational level. Here, the
flux of individual lines has to be derived, which requires that the selected
lines are isolated. Typically, this is not the case for Q transitions, which
are characterised by a constant total angular momentum (see
Fig.~\ref{fig:ohband}). Furthermore, the separation of variability groups
becomes even more difficult if the spectral resolution is relatively low.

To overcome these limitations, \skc{} uses a completely different approach to
obtain the scaling factors for the different line groups defined in
Sect.~\ref{sec:airglow}. In \skc{}, the contributions of the line groups to
each pixel of the sky spectrum are estimated. Subsequently, the resulting
spectra for each line class are scaled.

This is performed by applying the airglow model presented in the previous
section. The wavelengths and intensities of the lines and their group
identifications can be converted into intensities of the different line groups
for each pixel. This requires a convolution of the lines from the line list
with a kernel similar to the instrumental profile of the observed spectra. The
mean FWHM of the sky lines, which was obtained in a previous step (see
Sect.~\ref{sec:FWHM}), is used to create a sufficiently realistic Gaussian
kernel. In order to treat intensity ratios of overlapping lines as
realistically as possible, the airglow variability model from Noll et
al. (\cite{NOL12}) (see Sect.~\ref{sec:airglow}) was included. This allows one
to rougly correct for the influence of solar activity, season, and time of
night on the main line variability classes green O\,I, Na\,I\,D, red O\,I, OH,
and O$_2$.

Different line groups contributing to the same pixel implies that the sky
scaling factors cannot be derived by a simple division of line fluxes of
science and sky spectrum anymore. Instead, each scaling factor of the
individual line groups has to be included in a fitting procedure as a free
fitting parameter. For this purpose, the C version of the least-squares fitting
library MPFIT by C.~Markwardt\footnote{
\texttt{http://www.physics.wisc.edu/$\sim$craigm/idl/cmpfit.html}} based on the
FORTRAN fitting routine MINPACK-1 by Mor\'e et al. (\cite{MOR80}) is used. The
$\chi^2$ minimisation procedure of this routine is based on a
Levenberg-Marquardt technique, an iterative search algorithm characterised by
gradient-controlled jumps in parameter space. Since this technique is
potentially prone to find local minima, reasonable start values and constraints
on the fit parameters are required. For this reason, the mean ratios of the
line peaks in the science and sky spectrum are determined for each line group
(see Sect.~\ref{sec:airglow}). Only those spectrum pixels are included that
were identified as line peaks (see Sect.~\ref{sec:linesearch}) or are separated
from peaks by not more than half the line FWHM (see Sect.~\ref{sec:FWHM}) and
have a relative contribution of the selected line group of at least
{\sc weightlim} (default: 0.67; see Table~\ref{tab:setup}). Moreover, pixels
with unreasonable flux ratios are rejected by applying a global $\sigma$ limit
that is derived from the full set of line peaks and is provided by the
parameter {\sc siglim} (default: 15; see Table~\ref{tab:setup}). In this way,
strong object emission lines can be identified in the science line spectrum and
excluded. Finally, the $\sigma$-clipping approach with variable $\sigma$ limit
described in Sect.~\ref{sec:FWHM} is applied to the selected pixels of each
group separately in order to further improve the pixel selection. The
remaining pixels of this procedure are taken for the initial line group scaling
and fitting algorithm, i.e. only those pixels are considered for the $\chi^2$
calculation. If suitable pixels cannot be found for a line group, a mean flux
ratio of the corresponding system of roto-vibrational bands (e.g. OH; see
Sect.~\ref{sec:airglow}) or a global flux ratio is taken for $A$~groups and a
value of 1 is assumed for $B$~groups. For most sky spectra, this approach
should result in a good first guess sufficient for achieving a rapid
convergence to the global minimum.

As an option, the fitting can be restricted to uncertain line groups only. The
decision on the group selection depends on the parameter {\sc fitlim} (see
Table~\ref{tab:setup}) which provides a limiting ratio of the root mean square
(RMS) and the mean of the group-specific scaling factors. By default, this
value is set to 0, i.e. all fittable line groups are considered.

\subsection{Correction of wavelength grid}\label{sec:wavegrid}

Since the sky lines of the science spectrum are removed by a scaled reference
sky line spectrum, it is imperative that the wavelength grids of both spectra
are aligned. Differences of less than a pixel can already significantly
deteriorate the quality of the sky subtraction. Relatively large deviations can
occur if a lamp spectrum taken in daytime at different ambient conditions as
the science spectrum is used for the wavelength calibration. However, even
subpixel shifts that are routinely observed in data taken under perfect
conditions can cause problems.

For this reason, \skc{} offers an optional correction of the wavelength grid by
applying a Chebyshev polynomial of degree
$n_\mathrm{w}$
\begin{equation}
\lambda' = \sum_{i = 0}^{n_\mathrm{w}} c_i t_i,
\end{equation}
where
\begin{equation}
t_i = \left\{ \begin{array}{ll}
1 & \textrm{for\ } i = 0 \\
\lambda & \textrm{for\ } i = 1 \\
2 \, \lambda \, t_{i-1} - t_{i-2} & \textrm{for\ } i \ge 2
\end{array} \right.
\end{equation}
and $\lambda$ ranging from -1 to 1. The temporary conversion of the wavelength
grid to a fixed interval results in coefficients $c_i$ independent of the
wavelength range and step size of the input spectrum. The wavelength solution
is not changed if $c_1 = 1$ and $c_i = 0$ for all other $i$. It is possible
to set an individual start value for the constant term $c_0$ via the parameter
{\sc cheby\_const} (see Table~\ref{tab:setup}). In this way, significant
possible shifts between the wavelength grids of the science and the sky
spectrum can be considered.

The coefficients $c_i$ are determined by an iterative procedure. This process
is initialised with two subsequent estimates (for a better $\sigma$-clipping)
and a fit of the line flux correction factors (see Sect.~\ref{sec:linefit}).
During this first iteration the wavelength grid remains untouched. In the next
step, the coefficients $c_0$ and $c_1$ are fitted using MPFIT. Now, a new
estimate is calculated and the line flux correction factors are fitted again.
Then, the next iteration starts by fitting the wavelength grid, now applying a
Chebyshev polynomial of degree 2. After that, the line scaling factors are
adapted again in order to incorporate the change of the wavelength grid. Each
iteration increases $n_\mathrm{w}$ by 1 and uses the results of the
previous iteration as input. The search for the best polynomial degree is
controlled by the three input parameters {\sc cheby\_min}, {\sc cheby\_max},
and {\sc wtol} (see Table~\ref{tab:setup}). The iteration process is stopped
once the maximum polynomial degree given by {\sc cheby\_max} is reached. For a
value of -1, no wavelength grid correction is performed. The parameter
{\sc cheby\_min} indicates the minimum degree, i.e. the minimum number of
iterations. For $n_\mathrm{w}$ not less than {\sc cheby\_min}, the code checks
whether the resulting $\chi^2$ shows a relative $\chi^2$ improvement of at
least {\sc wtol} (default: $1 \times 10^{-3}$) compared to the best $\chi^2$, so
far. If this is not the case, the procedure stops and the results for the
polynomial with the lowest $\chi^2$ are taken. An exception is when
{\sc cheby\_min}~>~{\sc cheby\_max} is chosen. In this case, the code runs
until {\sc cheby\_max} is reached and the corresponding results for this degree
are taken, regardless of the results for the lower polynomial degrees. The
default values for {\sc cheby\_max} and {\sc cheby\_min} are 7 and 3,
respectively.

Independent of the use of a Chebyshev polynomial, the modified sky spectrum has
to be rebinned to the wavelength grid of the science spectrum. For this task,
the code offers two options, which can be selected by the parameter
{\sc rebintype} (see Table~\ref{tab:setup}). The first method adds up the
fractional fluxes of input pixels contributing to the wavelength range of the
output pixel. The second approach is based on the convolution of the input
spectrum with a pixel-dependent asymmetric damped sinc kernel
\begin{equation}
f(k) = e^{-((k - s) / \delta)^2} \, \frac{\sin(\pi (k - s))}{\pi (k - s)},
\end{equation}
with $k$ being an integer variable ranging from $-k_\mathrm{max}$ to
$k_\mathrm{max}$. The damping constant $\delta$ and the kernel radius
$k_\mathrm{max}$ are fixed and have the values 3.25 and 5. The parameter $s$ is
the subpixel shift of the sky spectrum relative to the science spectrum. It is
a function of the pixel position and ranges from $-0.5$ to $0.5$. For shifts
above half a pixel, complete pixels are treated by a simple renumbering of the
input pixels in the output spectrum. No convolution is performed for this
integer part of the pixel shift. The approach is similar to the one used in the
IDL routine {\tt sshift2d.pro} of the Lowell Buie Library\footnote{
\texttt{http://www.astro.washington.edu/docs/idl/htmlhelp/ slibrary30.html}}.
However, the original programme is only for a constant shift of the entire
spectrum. A wavelength-dependent shift is also possible, as long as the shift
changes slowly with the spectrum pixels and the pixel size is nearly constant
for the whole input and output wavelength grids. These requirements are
sufficiently met if inconsistencies of the wavelength grids are in the order of
1\,pixel and if the functional dependence of the differences can be described
by a low order polynomial. The relatively complicated rebinning method
described above is able to effectively suppress the broadening of spectral
lines, which typically occurs if a spectrum is rebinned to a shifted grid of
similar pixel size. The line-broadening suppression is achieved by alternating
positive and negative contributions to the kernel as incorporated in the sinc
shift method. Therefore, the sinc shift method produces the best results if
significant subpixel shifts close to half a pixel are frequent. However, for a
very good agreement of the wavelength grids with subpixel shifts close to zero,
it might be better to use the simple rebinning method. In such a case, the
relatively broad sinc kernel influences the spectrum more than simple
regridding.

\section{Test data set}\label{sec:data}

\begin{table}
\caption[]{Properties of the \xshoot{} test data set}
\label{tab:dataset}
\centering
\footnotesize
\vspace{5pt}
\begin{tabular}{l c c c}
\hline\hline
\noalign{\smallskip}
Par. & Set~1 & Set~2 & Set~3 \\
\noalign{\smallskip}
\hline
\noalign{\smallskip}
ESO ID
& 288.D-5015    & 088.A-0725    & 086.A-0974      \\
$N_\mathrm{exp}${}$^\mathrm{a}$
& $12,24,80$    & $4,4,4$       & $7,7,7$         \\
Dates [UT]
& 2011/12/25    & 2011/12/19    & 2010/12/29$-$31 \\
Times$^\mathrm{b}$ [UT]
& 00:58$-$03:25 & 03:12$-$04:02 & 05:56$-$07:31   \\
Slit width$^\mathrm{a}$ ['']
& $1.0,0.9,0.9$ & $1.0,0.9,0.9$ & $1.0,0.9,0.9$   \\
$T_\mathrm{exp}${}$^\mathrm{a}$ [s]
& $606,294,100$ & $705,609,246$ & $900,900,900$   \\
\noalign{\smallskip}
\hline
\end{tabular}
\footnotesize
\begin{list}{}{}
\item[$^\mathrm{a}$] The number of exposures $N_\mathrm{exp}$, the slit width, and
the exposure time $T_\mathrm{exp}$ are given for the three \xshoot{} arms UVB,
VIS, and NIR separated by commas.
\item[$^\mathrm{b}$] The time range was derived from the start of the first
exposure and the end of the last exposure of all three arms. For Set~3, it was
neglected that the observations had been performed in three different nights.
\end{list}
\end{table}

\begin{figure*}
\centering
\includegraphics[width=15.2cm,clip=true]{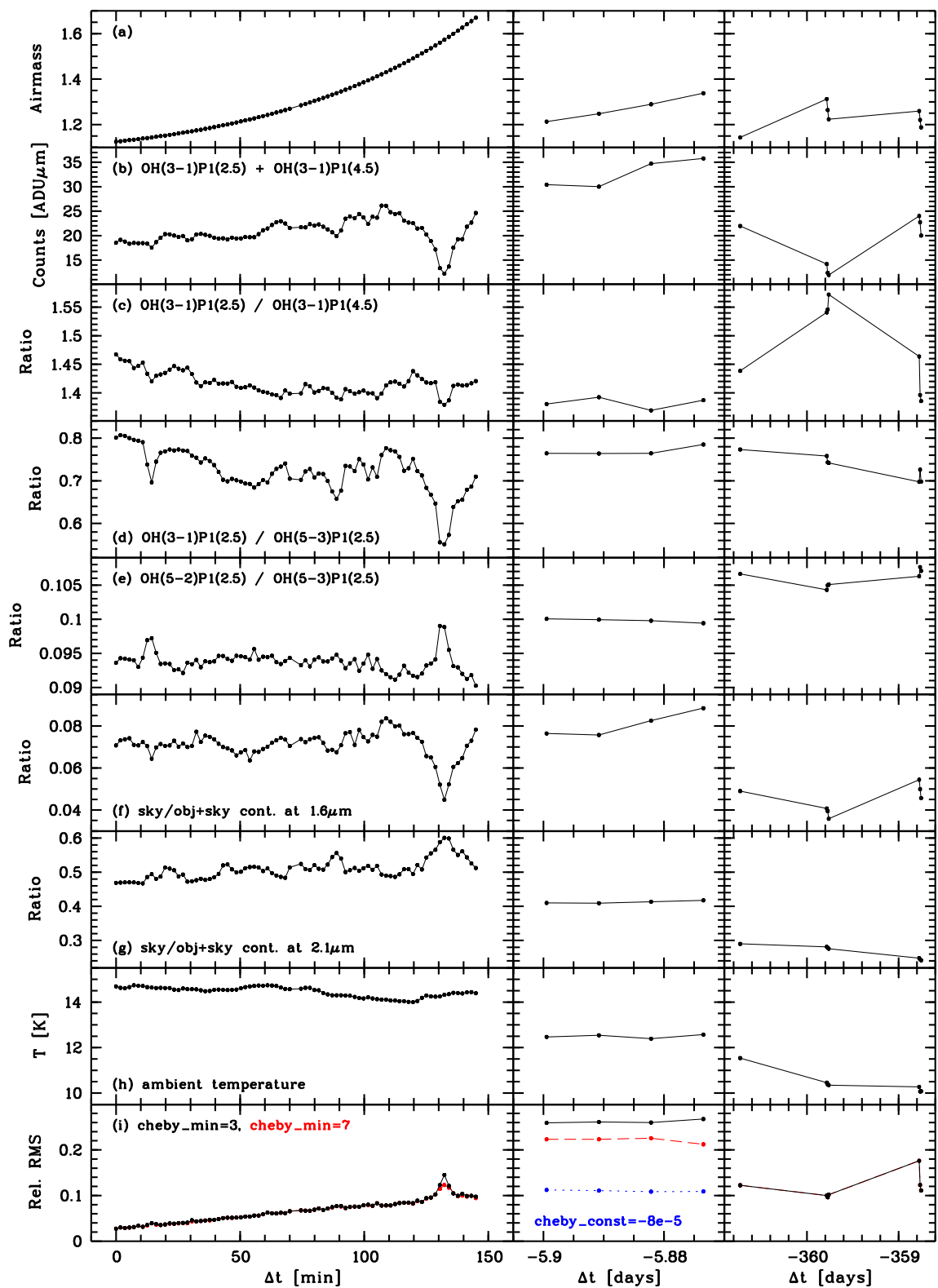}
\caption[]{Properties of the \xshoot{} NIR-arm test data and quality of the sky
subtraction by \skc{} as a function of time. The time difference $\Delta t$
is given relative to the object spectrum to be corrected for sky emission
(2011/12/25 00:59:05 UT). The left, middle, and right panels characterise the
sky data of the same night, about six nights before, and about one year before
the object reference exposure, respectively. From top to bottom, the panels
display (a) the airmass, (b) the summed intensity of two airglow lines of the
\mbox{OH(3-1)} band, (c) the ratio of the same lines tracing relative
$B$-group variability, (d) the ratio of the same rotational line in two
different OH bands tracing the relative $A$-group variability, (e) the ratio of
two OH lines which can only vary by effects not related to airglow, (f) the sky
continuum at 1.6\,$\mu$m relative to the corresponding measurement in the
uncorrected object spectrum, (g) a similar ratio for 2.1\,$\mu$m, (h) the
ambient temperature in K, and (i) the RMS of the sky subtraction residuals for
line pixels relative to the mean line peak flux. The latter is shown for two
different minimum degrees of the Chebyshev polynomial for the improvement of
the wavelength solution by \skc{}. In the middle panel of (i), the effect of an
initial shift of the wavelength scale by one pixel is also plotted (dotted
line).}
\label{fig:dataprop}
\end{figure*}

To illustrate the performance of \skc{}, we have used data taken with the VLT
echelle spectrograph \xshoot{} (Vernet et al. \cite{VER11}). This instrument is
well suited for testing sky subtraction, since its three arms UVB, VIS, and NIR
cover the entire wavelength range affected by significant airglow emission. In
fact, wavelengths from 0.3 to 2.5\,$\mu$m can be observed simultaneously with
medium resolution between 3300 and 18200 depending on the arm and the selected
slit widths. Since \xshoot{} is a slit spectrograph\footnote{The IFU of
\xshoot{} is an image slicer, which converts a 4''$\times$1.8'' field into a
12''$\times$0.6'' pseudo slit.}, the performance of \skc{} can be compared to
the classical sky subtraction method based on the interpolation of 2D data.

We have also investigated the \skc{} performance for data taken with other
instruments. The \skc{} User Manual that is provided along with the code also
shows several VLT FORS and SINFONI examples.

\subsection{Sample selection and data reduction}\label{sec:sampsel}

In order to analyse the sky subtraction quality as a function of time, we
searched for a time series of a single target in the ESO archive. We decided
to use the observations of the ESO programme 288.D-5015, which was carried out
on 25 Dec 2011 between 0:58 and 3:25 UT and comprises 12 UVB, 24 VIS, and 80
NIR-arm spectra with exposure times of 606, 294, and 100\,s, respectively (see
Set~1 in Table~\ref{tab:dataset}). The observed object was the ultracool white
dwarf SDSS\,0138-0016 (see Parsons et al. \cite{PAR12}). The spectrum is
characterised by a maximum emission at about 1\,$\mu$m, absorption bands in the
near-IR, and also emission lines in the optical. As reference object spectrum
to be corrected by \skc{}, we defined the first observation in each arm. For
all other observations, only the sky emission was used.

The \xshoot{} data were reduced using the ESO public pipeline release V2.0.0
executed with the Reflex workflow V2.3 (see Modigliani et al. \cite{MOD10}).
To obtain the sky subtraction results for the 2D interpolation method, we
directly used the output pipeline 1D spectra. For the correction with \skc{},
we reduced the data without sky subtraction and extracted the input spectra for
\skc{} from the resulting 2D spectra. For the object spectrum, we applied a
wavelength-independent trace based on the median object flux in spatial
direction. Pixels masked by the pipeline were excluded from the extraction and
the summed object flux was scaled to correct for missing pixels using the
object profile. This approach is similar to the method used by the pipeline.
However, by doing it ourselves, we know the exact sky area that contributes to
each wavelength. This information is important for extracting the sky from the
2D spectra. Here, we computed the median along the spatial direction and scaled
the resulting spectrum to have the same sky area as the reference object
spectrum. The data set was not flux calibrated, since the instrument response
and the atmospheric extinction (the latter at least for the VIS and NIR arm)
only slowly vary with wavelength and time compared to the defined airglow
variability groups (see Sect.~\ref{sec:airglow}).

In order to test \skc{} for sky data not taken in the same night as the
reference object spectrum, we extended our sample by two additional data sets
(see Set~2 and 3 in Table~\ref{tab:dataset}). The observations had the same
slit widths and long exposure times to avoid that the extracted median sky
would be contaminated by a bright object. Set~2 was taken six nights earlier
than the reference observations and comprises four spectra in each arm. Set~3
was taken about one year earlier in three subsequent nights and consists of
seven exposures in each arm. Since the exposure times deviated from those of
the reference data, the resulting 1D sky spectra were corrected by the exposure
time ratio.

\subsection{Sample properties}\label{sec:sampprop}

\begin{figure*}
\centering
\includegraphics[width=\textwidth,clip=true]{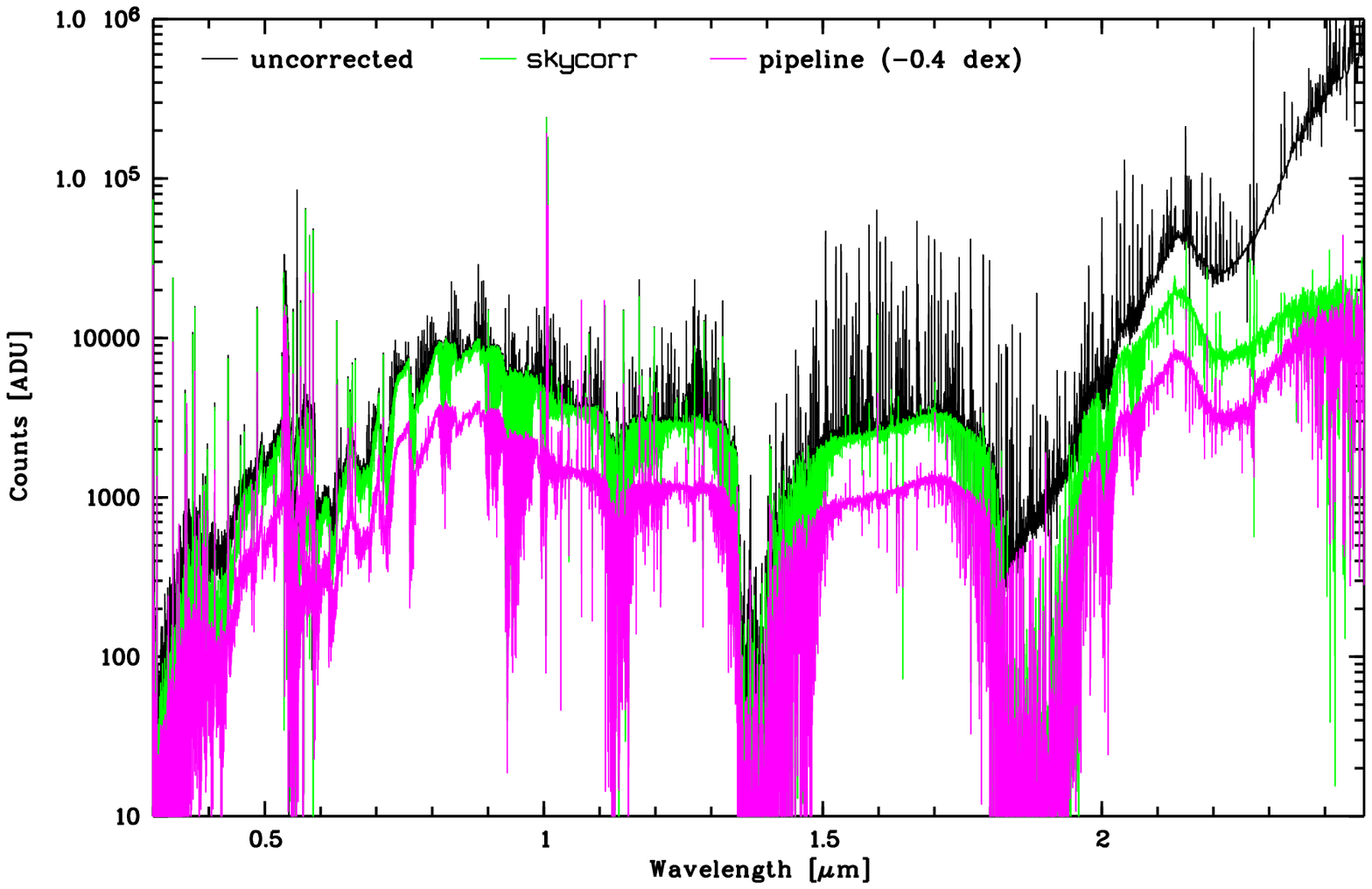}
\caption[]{Sky subtraction results for the three \xshoot{} arms for the sky
spectrum derived from the same 2D spectrum as the object spectrum
($\Delta t = 0$). The input object spectrum (black), the \skc{}-corrected
spectrum (green), and the corresponding \xshoot{} pipeline product based on
sky interpolation in the 2D spectrum (magenta) are shown. For a better
comparison, the latter is shifted by 0.4\,dex to lower values. The spectra are
not flux calibrated. This explains the apparent flux jump between the UVB and
VIS arm at about 0.59\,$\mu$m and the bump at about 2.14\,$\mu$m (cf. Parsons
et al. \cite{PAR12}). The strong uncorrected lines in the near-IR (mainly
between 1.0 and 1.35\,$\mu$m) were mostly caused by instrumental defects.}
\label{fig:refspec}
\end{figure*}

\begin{figure*}
\centering
\includegraphics[width=\textwidth,clip=true]{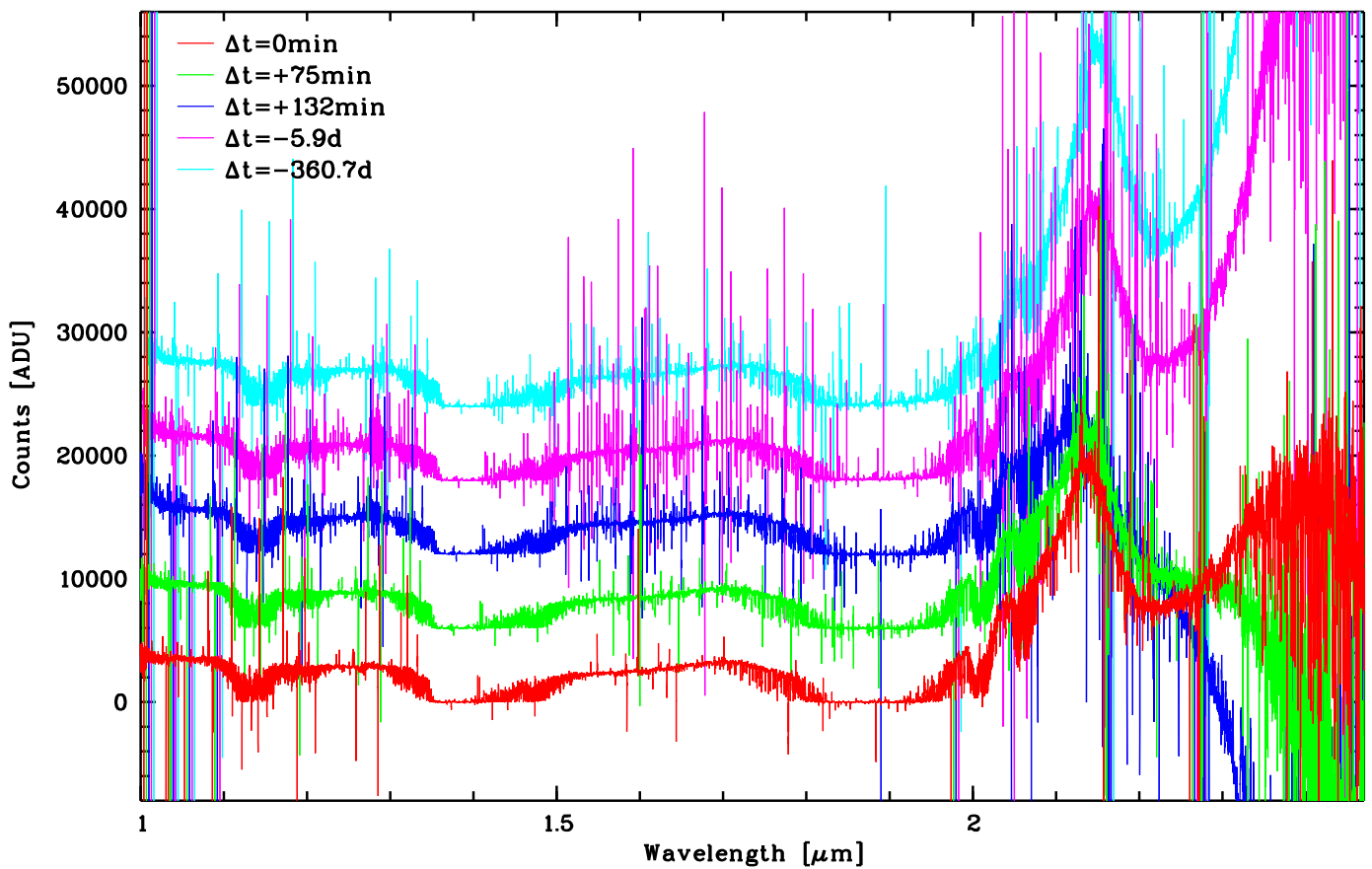}
\caption[]{\Skc{} output spectra for different \xshoot{} NIR-arm data sets
consisting of the same input object spectrum but different sky spectra. The
colour codes for the different time differences are given by the figure
legend. For a better comparison, adjacent spectra are shifted by 6000\,ADU and
3\,nm.}
\label{fig:compspec}
\end{figure*}

Figure~\ref{fig:dataprop} provides an overview of the observing conditions for
the investigated spectra. The data for each of the three subsamples (see
Table~\ref{tab:dataset}) are displayed in separate panels as a function of the
time interval between the mid-exposures of the sky and object observations.

The first row of panels (a) shows the change in airmass. The observation of a
single target over 2.5\,h in the main run results in a significant change in
the airmass. This affects the airglow intensity due to the nearly proportional
increase of the projected emission layer width (van Rhijn \cite{RHI21}; Noll et
al. \cite{NOL12}).

The summed intensity of two P-branch lines of the \mbox{OH(3-1)} band is shown
in the next row of panels. Apart from the airmass-related van Rhijn effect, the
intensity varies on different time scales due to changes of the chemical
composition, temperature, and dynamics in the OH emission layer (see Khomich et
al. \cite{KHO08}). For the entire data set, a maximum ratio of 3 in the
intensity can be observed, which illustrates the challenge in correcting the
sky in a science spectrum with data taken at a different time. The variations
are not only caused by airglow variability. Changes of the instrument
performance and calibration data can also have an effect. Moreover, variations
in the atmospheric transparency can be important for non-photometric
observations. This explains the dip in the intensity about 135\,min after the
start of the main run, when a small cloud probably covered the sky in the line
of sight\footnote{This interpretation is consistent with data from ESO's
Ambient Conditions Database.}. Note that the decrease of the airglow intensity
was moderate compared to the almost complete extinction of the object light.
The difference is caused by the fact that the observed star is a point source,
whereas airglow emission covers the entire sky. This allows a partial
compensation of extinction losses by scattering of light from other sky
positions into the line of sight (see Chamberlain \cite{CHA61}; Noll et al.
\cite{NOL12}). Running \skc{} with a sky spectrum taken under such unfavourable
conditions is a good test of the applicability of the method in challenging
situations.

Panels (c) to (e) of Fig.~\ref{fig:dataprop} show the time dependence of
different line ratios. They indicate a weaker variability than the intensity.
However, variations in the order of 10\% for OH can still be critical for a
reasonable correction of strong airglow lines. For this reason, the line groups
described in Sect.~\ref{sec:airglow} have been introduced. While (c) shows an
example of the time-dependent deviations between different $B$~groups, (d) and
(e) are examples of deviations in the intensity ratios of $A$~groups. Since
the bands in (e) have the same upper vibrational level $v' = 5$, the ratio
should be constant if only the airglow variability is considered. Therefore,
the significant changes (especially if different observing runs are compared)
trace the influence of instrumental effects and atmospheric transparency.

Panels (f) and (g) show the ratio of the sky and object+sky continuum for the
wavelengths 1.6 and 2.1\,$\mu$m, respectively. Since the sky continuum cannot
be adapted by \skc{} (see Sect.~\ref{sec:method}), the variation of such ratios
is critical for the quality of the continuum correction. In the $H$ band, the
sky continuum is weak. Except for the cloud event and the observations that
were taken about one year before the object spectrum, the variation of the sky
relative to the object is in the order of 1\% only. Hence, continuum errors by
the sky subtraction in the $H$ band should be small as long as the observed
object is not distinctly fainter than the investigated white dwarf. In the $K$
band, the thermal emission by the telescope is a strong component. For our test
data, the contribution is up to 60\% at 2.1\,$\mu$m, i.e. the telescope
emission is about as strong as the object emission. The flux ratio in (g) is
well correlated with the ambient temperature in (h), which determines the
mirror temperature. The temperature deviated by up to 4.6\,K from the value of
14.7\,K for the reference exposure. Due to the strong variability, a good
thermal continuum subtraction appears to be difficult to achieve in this
wavelength regime if an object is not significantly brighter than the thermal
emission, and/or the temperature differences between the object and sky
exposures are not distinctly lower than 1\,K. At least in the case of
\xshoot{}, the strong variations in the $K$ band could also be partly related
to instabilities of the flat-field lamps (J. Vernet, priv. comm.). Hence, the
continuum subtraction for Set~2 and 3 could be improved by using the same
flat-field as for Set~1. For flux-calibrated spectra, which we did not analyse,
the use of the same flat-field during the reduction of the science and standard
star spectra is the default setting in the \xshoot{} pipeline Reflex workflow.

\section{Results}\label{sec:evaluation}

In the following, we evaluate the performance of \skc{} on the data set
discussed in Sect.~\ref{sec:data}. At first, the test approach is described
(Sect.~\ref{sec:approach}). Then, the performance of \skc{} is compared with
the 2D sky interpolation method applied by the \xshoot{} pipeline
(Sect.~\ref{sec:deltatis0}). This test can only be carried out if there is no
time difference between the science and the sky spectra. Next, the performance
is analysed for different time intervals by using the entire test data set
(Sect.~\ref{sec:timediff}). Finally, we compare the results of \skc{} with
those obtained without line scaling (Sect.~\ref{sec:noscaling}) and using the
method of Davies (\cite{DAV07}) for our test data (Sect.~\ref{sec:davies}).

\subsection{Test approach}\label{sec:approach}

For the test runs, we used a fixed set-up of the parameters listed in
Table~\ref{tab:setup}. Since \skc{} was optimised to minimise user
interaction, most parameter values agree with the listed default values. The
parameter {\sc vac\_air} was set to `air' because the \xshoot{} pipeline
provides wavelengths in air. Since the \xshoot{} spectra have a
sufficiently high resolution, some parameters can be modified to speed up the
code without deteriorating the results. Setting {\sc fluxlim}~=~0.005 switches
off the iterative search of continuum windows between the lines (see
Sect.~\ref{sec:linesearch}) without changing the result. We also modified the
FWHM and $\chi^2$ convergence criteria {\sc ltol} ($1 \times 10^{-1}$) and
{\sc ftol} ($1 \times 10^{-2}$), which did not negatively affect the sky
subtraction quality. For \xshoot{} NIR-arm spectra with about 25000 pixels, the
run times were measured to be between 12 and 25\,s\footnote{run on a Core2Quad
Q9550@2.83GHz, 8GB RAM, Fedora 19 (64 bit)} mainly depending on the number of
calculated polynomials of different degrees (see Sect.~\ref{sec:wavegrid}).
Without changing the parameters discussed above, the run times would have been
between 12 and 36\,s. Since \xshoot{} spectra cover a wide wavelength range,
characterised by a roughly linear increase of the width of the line profile in
pixels, we set {\sc varfwhm} to 1 (see Sect.~\ref{sec:FWHM}). Finally, we
reduced {\sc siglim} (Sect.~\ref{sec:linefit}) from the default value 15 to 5,
since the test spectra show many outliers due to bad pixels and not all of them
were masked by the pipeline.

\Skc{} produces an output ASCII file which includes the best-fit parameters and
several other quantities that can be used to evaluate the sky subtraction
quality. In particular, the file provides the RMS of the error-weighted sky
subtraction residuals (no continuum) for line pixels relative to the
error-weighted mean line peak flux. The weights depend on the statistical noise
and possible systematic errors by bad pixels and similar defects. The selection
of line pixels originates from the line--continuum separation discussed in
Sect.~\ref{sec:linesearch}. The relative RMS, as defined above, has turned out
to be a good indicator of the sky line subtraction quality and will be used in
the following. Good values are in the order of a few per cent.

\subsection{Comparison with the 2D sky interpolation method}
\label{sec:deltatis0}

As a first test of the quality of the sky subtraction by \skc{}, we compare the
results for $\Delta t = 0$, i.e. the science and sky spectrum taken from the
same 2D spectrum, with those of the \xshoot{} pipeline.
Figure~\ref{fig:refspec} shows these resulting spectra in the three \xshoot{}
arms. Both methods produce convincing spectra of the target star (cf. Parsons
et al. \cite{PAR12}). In the UVB arm ($0.30 - 0.59$\,$\mu$m), stellar emission
lines can be seen. The VIS arm ($0.53 - 1.02$\,$\mu$m) mainly reveals stellar
absorption bands and the NIR arm ($0.99 - 2.48$\,$\mu$m) shows strong
atmospheric absorption of the stellar continuum (essentially by water vapour
bands). Note that the spectra are not flux calibrated (see
Sect.~\ref{sec:sampsel}). The UVB and VIS-arm spectra are only affected by
relatively weak sky features compared to the object continuum (cf.
Sect.~\ref{sec:airglow}). Therefore, a good sky correction is not particularly
difficult for the given example. The situation is much more challenging in the
\xshoot{} NIR-arm range, where the sky emission tends to be significantly
brighter, reaching up to 1 to 2 orders of magnitude higher intensities than the
stellar continuum. For this reason, we will focus the tests in
Sect.~\ref{sec:evaluation} on the NIR arm.

Despite the strong airglow lines, \skc{} and the 2D interpolation method
produce sky-corrected spectra with relatively smooth object continua. The
strongest residuals can be identified as instrumental defects (bad pixels). The
most obvious difference between the results of both methods is the deviating
continua at the red margin of the NIR arm. This discrepancy can be explained by
the very strong thermal background, intensity gradients along the spatial
direction in the 2D spectra, and the differences in the extraction of the 1D
spectra (see Sect.~\ref{sec:sampsel}). Thus, the deviations are not related to
the sky subtraction quality. For a more quantitative comparison of this
quality, we have calculated the relative RMS defined in
Sect.~\ref{sec:approach}. Interestingly, the \skc{} output NIR-arm spectrum
has a significantly smaller RMS (0.027) than the corresponding pipeline
spectrum (0.040). This result is robust. Modifying the considered wavelength
range or the exclusion algorithm for unreliable RMS outliers (bad pixels) did
not change it significantly. Even though this is just one example and the
approach for the extraction of the 1D spectra could influence the RMS, \skc{}
appears to be able to subtract sky emission lines at least as well as a method
that benefits from the full 2D spectral information.

\subsection{Effect of time differences between science and sky spectra}
\label{sec:timediff}

\begin{figure}
\centering
\includegraphics[width=8.8cm,clip=true]{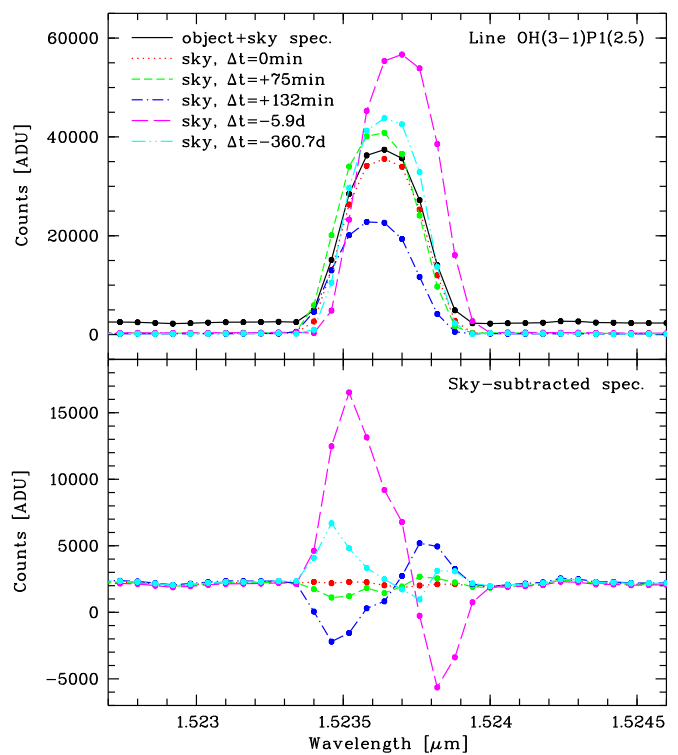}
\caption[]{Subtraction of the strong airglow line \mbox{OH(3-1)P1(2.5)} for
several time intervals between the object and the sky spectrum. The upper panel
shows the line in the input object (black) and sky spectra (see legend for
colour and line type). The lower panel displays the corresponding sky
subtraction residuals. To enhance their visibility, the ordinate was
significantly zoomed.}
\label{fig:skylinecorr}
\end{figure}

\begin{figure}
\centering
\includegraphics[width=8.8cm,clip=true]{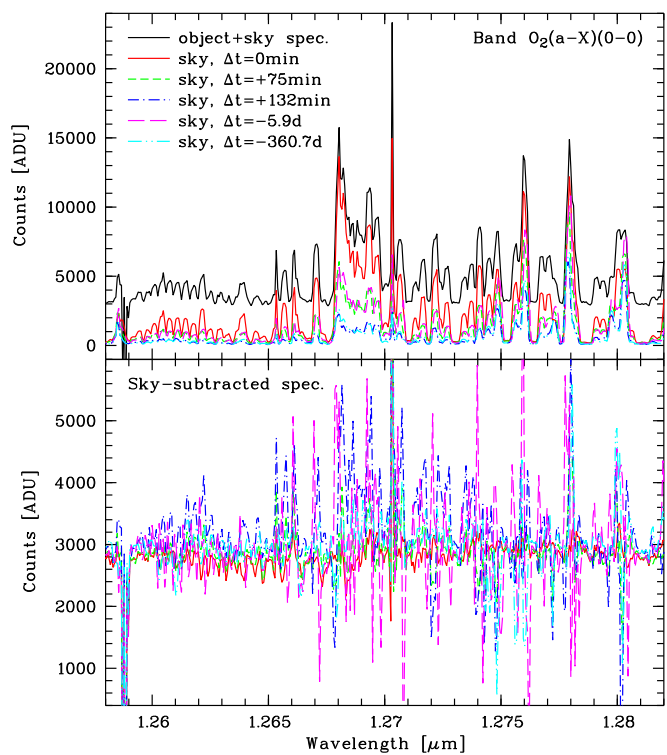}
\caption[]{Subtraction of the strong and only partly resolved airglow band
\mbox{O$_2$(a-X)(0-0)} (cf. Fig.~\ref{fig:o2a00band}) for several time
intervals between the object and the sky spectrum. The upper panel shows the
band in the input object (black) and sky spectra (see legend for colour and
line type). The lower panel displays the corresponding sky subtraction
residuals. Note that the ordinates are different in both panels. The two strong
uncorrected lines in the centre and the left margin of the figure were caused
by instrumental defects. Longwards of 1.275\,$\mu$m, the O$_2$ lines are
blended with \mbox{OH(8-5)} lines.}
\label{fig:skybandcorr}
\end{figure}

\begin{figure}
\centering
\includegraphics[width=8.8cm,clip=true]{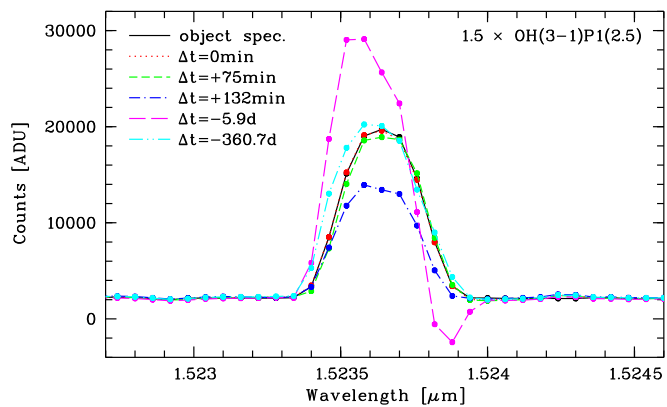}
\caption[]{Subtraction of the strong \mbox{OH(3-1)P1(2.5)} line from an object
spectrum with an artificial emission line (black) at the same position and
with the same line shape but only 50\% of the strength of the sky line. The
sky correction results are shown for several time intervals between the object
and the sky spectrum (see legend for colour and line type).}
\label{fig:addobjline}
\end{figure}

The main purpose of \skc{} is the subtraction of the sky emission in science
spectra by means of sky data taken at a different time and sky position. Using
the data set described in Sect.~\ref{sec:sampsel}, the quality of this
correction can be investigated as a function of time. In
Sect.~\ref{sec:sampprop}, we have discussed the sky variability during the
selected observing runs on the basis of the data plotted in
Fig.~\ref{fig:dataprop}. The last row of this figure shows the relative RMS
(see Sect.~\ref{sec:approach}) for the NIR-arm spectra of the test data set.
For the \skc{} parameter set-up listed in Table~\ref{tab:setup} (black
symbols and lines), the left-hand panel indicates a gradual increase of the
RMS from 0.027 at the beginning, over 0.067 at half of the time, to 0.098 at
the end of the run. A special situation occurred at $\Delta t \approx 132$\,min
with a maximum RMS of 0.145. As discussed in Sect.~\ref{sec:sampprop}, a cloud
probably covered the sky in the target direction. This caused a significant
decrease of the sky subtraction quality. However, these results are still
better than the sky correction performed with sky data taken six nights before.
Here, the relative RMS was about 0.26. Concerning the atmospheric conditions,
the most noticeable difference to the reference object exposure was the almost
two times higher OH emission intensity (middle panel in row (b)). This probably
contributes to the worse performance of \skc{}. However, a long time difference
does not inevitably mean a bad sky correction quality, as Set~3 in the
right-hand panel demonstrates. Although the time interval was about one year,
the RMS values are between 0.10 and 0.18, which is significantly better than
for $\Delta t = -6$\,d and even comparable with the end of the main run. The OH
intensity for Set~3 was closer to the reference value than for Set~2, but the
discrepancies are still quite large. Moreover, rows (d) to (f) of
Fig.~\ref{fig:dataprop} suggest significant differences in the instrumental
properties and calibration (see Sect.~\ref{sec:sampprop}).

In order to better understand the \skc{} performance differences,
Fig.~\ref{fig:compspec} shows the sky-subtracted NIR-arm spectra for five
different cases: $\Delta t = 0$\,min (the reference exposure), $+75$\,min (half
of the time of the main run), $+132$\,min (cloud), $-5.9$\,d (first exposure of
Set~2), and $-360.7$\,d (first exposure of Set~3). In addition,
Fig.~\ref{fig:skylinecorr} displays the input sky lines and the corresponding
sky subtraction residuals for the single strong airglow line
\mbox{OH(3-1)P1(2.5)} (see Sect.~\ref{sec:airglow}). Fig.~\ref{fig:skybandcorr}
shows a similar plot for the only partly resolved \mbox{O$_2$(a-X)(0-0)} band,
which is also blended with \mbox{OH(8-5)} lines. The figures confirm that the
strongest sky line residuals are found for the cloud event and Set~2. The
sky-corrected spectrum for the Set~3 example is an intermediate case. Further
implications are discussed in the following paragraphs.

Concerning the continuum correction, the cloud event changed the thermal
background in a way that the sky-corrected object continuum is completely wrong
in the red part of the $K$ band (Fig.~\ref{fig:compspec}). Significant changes
in the background can already be observed for $\Delta t = 75$\,min. Since the
sky continuum cannot be adapted to changing observing conditions, the continuum
correction especially in the red part of the $K$ band has to be taken with care
if the object is distinctly fainter than the background (see
Fig.~\ref{fig:refspec}).

As already indicated by Fig.~\ref{fig:dataprop}, the intensities of the OH
lines deviated the most from the reference exposure for the cloud event and
Set~2. Figure~\ref{fig:skylinecorr} also reveals that these cases show a
significant shift of the line centres. In particular, the Set~2 spectrum is
shifted by about one pixel. Since the physical changes of airglow line
positions are orders of magnitude too small to be visible in the \xshoot{}
data, the shifts are probably caused by instrumental effects related to e.g.
the position of the orders on the chip, variations in the wavelength
calibration frames, or uncertainties in the determination of the wavelength
solution by the pipeline. As the strong asymmetric residuals for the critical
cases suggest, the wavelength shifts appear at least as important for the sky
subtraction quality as the intensity of the lines. \Skc{} corrects the input
wavelength solution by fitting Chebyshev polynomials, where the degree is
increased in an iterative procedure (see Sect.~\ref{sec:wavegrid}). The default
algorithm is a search that at least checks degree 3 and in maximum degree 7
(see Table~\ref{tab:setup}). Interestingly, the critical cases were only
checked up to the minimum degree {\sc cheby\_min}~=~3. Therefore, we ran the
code for {\sc cheby\_min}~=~7, which significantly increased the code run time
but also improved the RMS for the cloud event and Set~2, as
Fig.~\ref{fig:dataprop} (red symbols and lines in last row) indicates.
Surprisingly, the best fit for the critical cases was achieved without changing
the input wavelength solution. Hence, the iterative procedure appeared to
mainly improve the fitting of the line intensities. The parameter
{\sc cheby\_const} (see Table~\ref{tab:setup}) allows a constant initial shift
of the wavelength grid. By setting it to $-8 \times 10^{-5}$, which corresponds
to about 1~pixel for \xshoot{} NIR-arm spectra, we can compensate the shift
found for the Set~2 example. The resulting RMS is also shown in
Fig.~\ref{fig:dataprop}. For the selected Set~2 spectrum, it decreased from
0.26 to 0.11. This striking improvement suggests that the fitting procedure of
the standard run did not find the global $\chi^2$ minimum due to strong
differences in the airglow intensities and wavelength grids of the input
science and sky spectra. In such critical cases, the \skc{} parameter set-up
can be optimised.

For blended O$_2$ lines, Fig.~\ref{fig:skybandcorr} indicates that the most
significant residuals are for the cloud case and Set~2 again. Nevertheless, the
situation is different, since the O$_2$ lines showed a completely different
time dependence compared to the OH lines. This can be seen at wavelengths
beyond 1.275\,$\mu$m, where both kinds of lines are present. At $\Delta t = 0$,
the O$_2$ band was distinctly brighter than for all other examples plotted. In
particular, the observations of Set~2 and 3 showed intensities that were lower
by a factor of 5 to 6. In view of the required strong intensity corrections and
the blending of lines with different variability, the quality of the recovered
stellar continuum is remarkably good. Only the spectrum related to the cloud
shows significant continuum offsets at the wavelengths where line blending is
most critical. It should be noted that the target spectrum retrieval can be
more difficult than in the present case if the observed object has a complex
spectrum in the range of the O$_2$ band. Then, the separation of blended sky
lines and continuum could be systematically wrong, which would be a problem if
the airglow lines need to be scaled by factors far from 1. The sky line
subtraction at about 1.27\,$\mu$m is certainly the most difficult for
\xshoot{}. For lower resolution spectra, additional wavelength ranges with
blended sky lines could have similar problems.

The NIR-arm spectrum of SDSS\,0138-0016 (see Fig.~\ref{fig:refspec}) does not
allow the test of how object emission lines would appear after the subtraction
of strong sky lines at similar wavelengths. For this reason, we carried out a
challenging test by multiplying the strong OH line in
Fig.~\ref{fig:skylinecorr} by 1.5 in the object spectrum and running \skc{}
with the full data set for this modified input spectrum. This modification
simulates an object emission line with the position, shape, and 50\% of the
intensity of the OH line. Fig.~\ref{fig:addobjline} shows the sky subtraction
results for the five selected cases compared with the expected object spectrum.
Apart from the $\Delta t = 0$ and 75\,min cases, which indicate an almost
perfect line reconstruction, the Set~3 spectrum taken almost one year before
the reference spectrum also shows a convincing line profile. This result
suggests that the conservation of object lines by means of the line grouping of
\skc{} (see Sect.~\ref{sec:airglow}) works in general. Unsatisfying line
profiles are only found for the two difficult cases. The quality of the sky
subtraction depends on the differences between the sky in the input science and
sky spectra as well as on the ratio of the object and sky line intensities.

\subsection{Comparison with sky subtraction without scaling}
\label{sec:noscaling}

\begin{figure}
\centering
\includegraphics[width=8.8cm,clip=true]{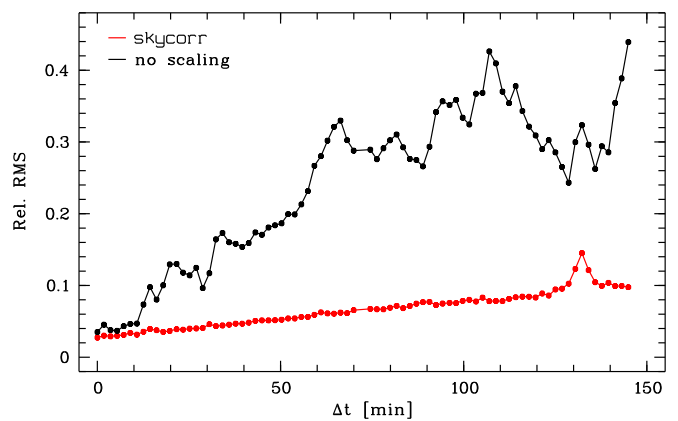}
\caption[]{Comparison of the RMS of the sky subtraction residuals relative to
the mean line peak flux (see Sect.~\ref{sec:approach}) by \skc{} (red) and
without line scaling (black) for the data set of the \xshoot{} NIR-arm sky
spectra taken in the same night as the corrected object spectrum. The results
are shown as a function of the time difference between the object and the sky
spectrum.}
\label{fig:noscalrms}
\end{figure}

Figure~\ref{fig:noscalrms} shows the resulting RMS of Set~1 for \skc{} (cf.
Fig.~\ref{fig:dataprop}) and the sky spectrum just subtracted from the science
spectrum without scaling. In the first 10\,min, the RMS was only slightly
higher than for \skc{} (ratios between 1.2 and 1.5). This indicates that for
time intervals of a few minutes, a significant correlation of the sky line
intensities can be assumed. Afterwards, the RMS ratio was quickly increasing,
reaching ratios up to 5.5. For Set~2 and 3, the RMS ratios ranged from 3.3 to
4.1 and from 2.4 to 5.1, respectively. These factors demonstrate that \skc{}
performs significantly better than a simple on-off technique without sky
scaling. They also show that the latter method is only reliable for very short
time differences.

\subsection{Comparison with Davies' method}\label{sec:davies}

\begin{figure}
\centering
\includegraphics[width=8.8cm,clip=true]{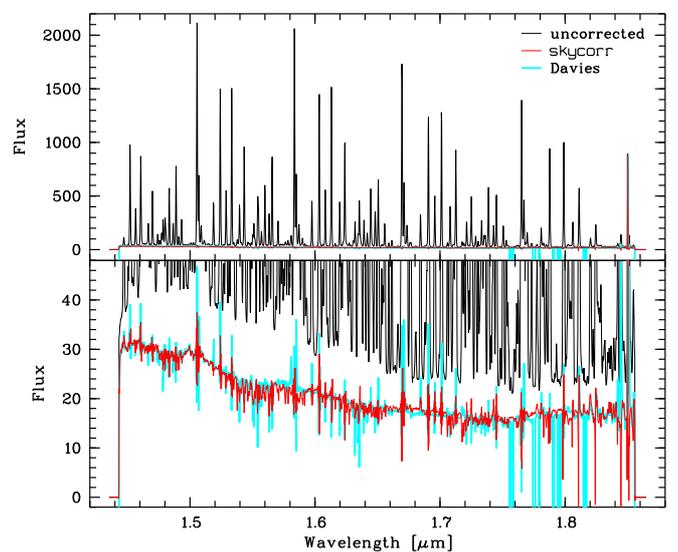}
\caption[]{Comparison of \skc{} (red) and Davies' code results (cyan) for an
example SINFONI $H$-band spectrum of the ESO observing programme 083.B-0456.
While the upper panel shows the full input science spectrum (black), the lower
panel focuses on the sky subtraction results. Highly negative fluxes (only
Davies' code) mark pixels with undefined fluxes. The time difference between
the object and sky exposures was 39\,min. Note that the faint object spectrum
was derived from the sky cube. The originally targeted object could not be
detected.}
\label{fig:sinfoH}
\end{figure}

\begin{figure}
\centering
\includegraphics[width=8.8cm,clip=true]{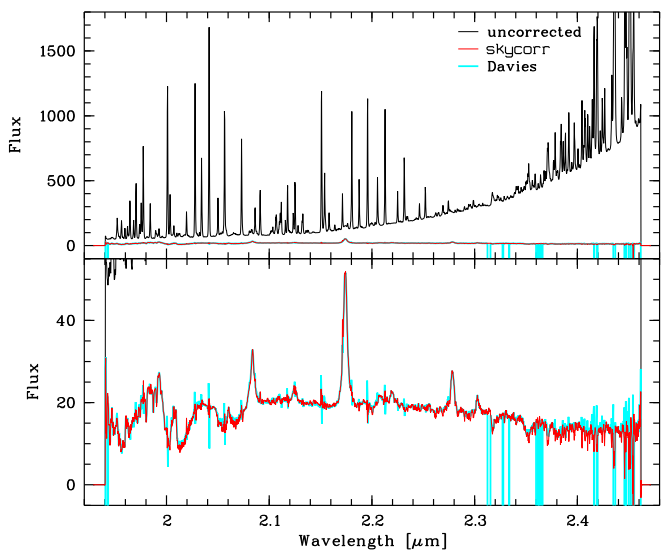}
\caption[]{Comparison of \skc{} (red) and Davies' code results (cyan) for an
example SINFONI $K$-band spectrum of the ESO observing programme 075.B-0648
showing the Seyfert\,2 galaxy NGC\,6240 (cf. Rosenberg et al. \cite{ROS12}).
While the upper panel shows the full input science spectrum (black), the lower
panel focuses on the sky subtraction results. Highly negative fluxes (only
Davies' code) mark pixels with undefined fluxes. The time difference between
the object and sky exposures was 11\,min.}
\label{fig:sinfoK}
\end{figure}

As far as we know, the method of Davies (\cite{DAV07}) is the only existing
approach with a similar philosophy and purpose as \skc{}. For this reason,
it is prudent to compare the performance of this method and \skc{}. To this
end, we attempted to run Davies' IDL code for the \xshoot{} NIR-arm test data
set defined in Sect.~\ref{sec:sampsel}. Since the version 1.1 implemented in
the currently released ESO SINFONI pipeline (Modgiliani et al. \cite{MOD07}) is
already fairly old, we improved the results by using the most recent version
2.0. Since SINFONI is an IFU spectrograph (Eisenhauer et al. \cite{EIS03}), 3D
data cubes with the spectra of a spatial 2D grid are required as input.
However, Davies' code also performs the sky line scaling on 1D spectra that are
derived from the filtered and integrated cubes for object and sky observations.
By replacing the internal 1D arrays for object and sky by the \xshoot{} input
data and writing the sky-subtracted object array into a file (neglecting the
subsequent application of the best solution to the object cube), we were able
to run Davies' code and to compare it with \skc{}. The runs were performed with
default parameters plus the thermal background subtraction and the wavelength
scale correction (via cross correlation) switched on.

The test indicates that \skc{} outperforms Davies' algorithm for \xshoot{}
data. The SINFONI-optimised Davies' code appears to be unstable for the
investigated data because of an erratically varying quality of the correction.
Possible reasons are the very wide wavelength range compared to SINFONI, the
higher resolution, the low S/N ratio, the numerous bad pixels, the use of
non-flux-calibrated spectra, and a complex continuum which cannot be fitted by
a single greybody. Despite the higher complexity, \skc{} was also faster than
the 1D part of Davies' code, which might be related to the use of C instead of
IDL.

For a fair comparison of \skc{} and Davies' method, we also have to consider
SINFONI spectra, for which the latter method was designed and works reliably.
In order to illustrate the performance of both codes for SINFONI data, we
studied an $H$-band (Fig.~\ref{fig:sinfoH}) and a $K$-band example
(Fig.~\ref{fig:sinfoK}). Since we are interested in the results for significant
time intervals, we did not use the sky cubes that belong to the object cubes.
Instead, we took the sky cubes from other observations of the same observing
programmes. In this way, time differences of 39 and 11\,min for the $H$ and $K$
band could be achieved. The data were reduced with ESO's SINFONI pipeline.
However, we used version 2.0 of Davies' code for the sky subtraction, as
already mentioned above. The code flags were the same as for the \xshoot{} data
set. For running \skc{} and the comparison in Figs.~\ref{fig:sinfoH} and
\ref{fig:sinfoK}, we used the 1D spectra produced by Davies' code.

For both examples, the two codes show a good performance, even though the
objects are up to two orders of magnitude fainter than the sky. Nevertheless,
the \skc{} sky line residuals are significantly weaker. For smaller time
intervals, the quality of the sky subtraction tends to be more similar, as
other tests indicated. The corrected object continua show only slight
deviations. For longer time intervals, this might change, as both codes use
different algorithms for the continuum correction (see Sects.~\ref{sec:intro}
and \ref{sec:contsub}).

In conclusion, the algorithms implemented in \skc{} seem to result in a
consistently better sky subtraction. Our code is more robust and flexible, i.e.
it can reliably be applied to data of different instruments and set-ups.

\section{Conclusions}\label{sec:conclusions}

We have presented a new method to subtract the sky in 1D spectra by means of
optimised plain sky spectra taken at different times and sky positions. This
method, which will be provided to the community as software package \skc{}, has
been inspired by the approach described by Davies (\cite{DAV07}) but it is more
sophisticated. Important features are an iterative separation of sky lines and
continuum, detailed line grouping based on an airglow model, pixel-based
scaling of line groups, and an adaptation of the wavelength solution of the
input sky spectra by Chebyshev polynomials and asymmetric damped sinc kernels.
The sky correction of an input science spectrum consists of the subtraction of
the best-fit scaled airglow lines and the separated continuum of the reference
sky spectrum.

We have tested the performance of the instrument-independent code by means of
\xshoot{} spectra. Fixing the object spectrum to be corrected, we have
analysed the sky subtraction quality for different time intervals between the
science and plain sky spectra. This comparison revealed promising results that
depend on the change of the airglow intensity, atmospheric transparency, and
the instrument calibration. The latter includes changes of the instrument
sensitivity and especially the wavelength calibration. Although the best
corrections are possible for time intervals of only a few minutes, where the
airglow intensities are still very similar, convincing sky corrections could
even be achieved for time differences of about one year. The \skc{} results
could also be compared with the 2D sky interpolation method of the \xshoot{}
pipeline, Davies' method, and the simple on-off method without sky fitting. In
conclusion, \skc{} performed at least as well as the 2D approach and
convincingly better than Davies' SINFONI-optimised code. The latter result
could be confirmed for two SINFONI examples and is due to the more complex sky
optimisation of \skc{}, the more robust performance, and the high flexibility
in terms of the properties of the input data. Compared to sky subtraction
without fitting, the \skc{} sky line residuals for the test data set without
parameter optimisation were between 2.1 and 5.5 times lower if short time
intervals of a few minutes are neglected.

These promising results suggest that \skc{} can be a valuable tool for
wavelength ranges where airglow emission lines dominate the sky background and
instrument set-ups which either offer only 1D data or require additional
observations if the 2D science exposure does not provide plain sky. Even for
sky subtraction in multi-object spectra without time differences, it might
be useful if instrument parameters, such as the slit width and the wavelength
coverage, are the same for object and sky spectra.

\Skc{} has been developed for Cerro Paranal. However, it is feasible to use it
for spectra taken at other observing sites. The specific Cerro Paranal
airglow variability model is mainly used for setting the initial weights for
the different line groups for each pixel. Since this model only considers five
classes, it could only have an influence if there was a significant
contribution of e.g. an OH and O$_2$ line to a certain pixel. This is
relatively rare. Moreover, pixels with a strong blending of line groups are
usually excluded from the fitting procedure. Ratios of lines of the same
molecule as provided by the model line list can also vary depending on the
observing site. However, these changes are distinctly weaker than the overall
intensity variations and hardly relevant for the weights. Finally, short-term
intensity variations can cause significant deviations from the airglow model.
Therefore, any intensity prediction is relatively rough, even for well studied
observing sites. In any case, the largest impact on the sky subtraction quality
is the reference sky spectra, which have to be taken for each instrumental
set-up individually.

\begin{acknowledgements}
We thank R. Davies for providing his IDL code. We are also grateful to the
anonymous referee for his/her detailed and valuable comments. This project made
use of the ESO archive facility. \xshoot{} and SINFONI data of the programmes
075.B-0648, 083.B-0456, 086.A-0974, 088.A-0725, and 288.D-5015 were analysed.
This study was carried out in the framework of the Austrian ESO In-Kind project
funded by BM:wf under contracts BMWF-10.490/0009-II/10/2009 and
BMWF-10.490/0008-II/3/2011. This publication is also supported by the Austrian
Science Fund (FWF). S. Noll receives funding from the FWF project P26130. W.
Kausch is also funded by the project IS538003 (Hochschulraumstrukturmittel)
provided by the Austrian Ministry for Research (bmwfw).
\end{acknowledgements}

\end{document}